\documentclass[aps,preprint,showpacs,superscriptaddress,groupedaddress]{revtex4}  
\usepackage{appendix}
\usepackage{graphicx}  
\usepackage{dcolumn}   
\usepackage{bm}        
\usepackage{amssymb}   
\usepackage{amsmath}
\usepackage{array}
\usepackage{epstopdf}
\usepackage{mathtools}
\usepackage[T1]{fontenc}  
\usepackage{textcomp}     
\usepackage{mathtools,slashed}
\usepackage{graphics}
\usepackage{epsfig}
\usepackage{psfrag}
\usepackage{color}
\usepackage[utf8]{inputenc}
\usepackage{scrextend}
\usepackage{setspace}
\DeclareMathAlphabet{\pazocal}{OMS}{zplm}{m}{n}
\usepackage{amsfonts}
\usepackage{amsthm}
\usepackage{float}
\usepackage{textcomp}
\usepackage[section]{placeins}
\usepackage[caption=false]{subfig}
\usepackage{subfig}
\usepackage{tabularx}
\usepackage{multirow}

\newcommand*{\rom}[1]

\begin{document}
\title{Observable Proton Decay in Flipped $\bm{SU(5)}$}
\date{\today}
\author{Maria Mehmood}
\email[E-mail: ]{mehmood.maria786@gmail.com}
\affiliation{Department of Physics, Quaid-i-Azam University, Islamabad 45320, Pakistan}
\author{Mansoor Ur Rehman}
\email[E-mail: ]{mansoor@qau.edu.pk}
\affiliation{Department of Physics, Quaid-i-Azam University, Islamabad 45320, Pakistan}
\author{Qaisar Shafi}
\email[E-mail: ]{shafi@bartol.udel.edu}
\affiliation{Bartol Research Institute, Department of Physics and Astronomy, University of Delaware, Newark, DE 19716, USA}

\begin{abstract}
We explore proton decay in a class of realistic supersymmetric flipped $SU(5)$ models supplemented by a $U(1)_R$ symmetry which plays an essential role in implementing hybrid inflation. Two distinct neutrino mass models, based on inverse seesaw and type I seesaw, are identified, with the latter arising from the breaking of $U(1)_R$ by nonrenormalizable superpotential terms. Depending on the neutrino mass model an appropriate set of intermediate scale color triplets from the Higgs superfields play a key role in proton decay channels that include $p \rightarrow (e^{+},\mu^+)\, \pi^0$, $p \rightarrow ( e^+,\mu^{+})\, K^0 $, $p \rightarrow \overline{\nu}\, \pi^{+}$, and $p \rightarrow \overline{\nu}\, K^+ $. We identify regions of the parameter space that yield proton lifetime estimates which are testable at Hyper-Kamiokande and other next generation experiments. We discuss how gauge coupling unification in the presence of intermediate scale particles is realized, and a $Z_4$ symmetry is utilized to show how such intermediate scales can arise in flipped $SU(5)$. Finally, we compare our predictions for proton decay with previous work based on $SU(5)$ and flipped $SU(5)$.
\end{abstract}
\pacs{12.10.-g,13.30.-a,12.60.Jv}
\maketitle
\section{Introduction}
Proton decay is rightly considered an important observable and discriminator for models of Grand Unified Theories (GUTs). The current lifetime bounds on various proton decay channels by Super-Kamiokande (Super-K) \cite{Miura:2016krn,Abe:2013lua,Takhistov:2016eqm,Regis:2012sn,Abe:2014mwa,Kobayashi:2005pe}, and the anticipated experimental results from the next generation experiments such as JUNO \cite{An:2015jdp}, DUNE \cite{Acciarri:2015uup}, and Hyper-Kamiokande (Hyper-K) \cite{Abe:2018uyc} should provide valuable information for comparing the proton decay predictions by GUT models. In this regard  proton decay induced by the dimension five operators in supersymmetric GUTs has been a subject of great interest. The expected dominant decay mode, $p \rightarrow K^+ \overline{\nu}$, in  minimal supersymmetric $SU(5)$ has been under  intense scrutiny \cite{Goto:1998qg,Murayama:2001ur}. This decay mode can be suppressed by assuming suitably large sfermion masses \cite{Hisano:2013exa,Nagata:2013sba,Nagata:2013ive,Evans:2015bxa,Ellis:2015rya,Ellis:2016tjc,Evans:2019oyw,Ellis:2019fwf} now also favored by LHC searches \cite{LHC}. This observation leads to an interesting comparison of gauge boson mediated proton decay modes in $SU(5)$ and flipped $SU(5)$ \cite{DeRujula:1980qc,Georgi:1980pw,Barr:1981qv,Derendinger:1983aj,Antoniadis:1987dx,Barr:1988yj,Shafi:1998dv} models via the dimension six operators as recently discussed in \cite{Ellis:2020qad}. The flipped $SU(5)$ model considered in  \cite{Ellis:2020qad} is based on a no-scale supergravity model of inflation with an approximate $Z_2$ symmetry and modified $R$ parity.

In a recent paper \cite{Lazarides:2020bgy} an exciting possibility of observable proton decay from a supersymmetric $SU(4)_c \times SU(2)_L \times SU(2)_R$ (4-2-2) model mediated by color triplets of intermediate mass range was identified. This 4-2-2 model nicely implements shifted hybrid inflation, as shown in \cite{Lazarides:2020zof}. These studies have prompted the present paper where we consider a supersymmetric hybrid inflation model \cite{Kyae:2005nv,Rehman:2009yj,Rehman:2018nsn} based on the flipped $SU(5)$ gauge symmetry, supplemented by a global $U(1)_R$ symmetry and $\mathbb{Z}_2$ matter parity. In order to study the contributions to proton decay from the color triplets of 5-plet and 10-plet Higgses, we employ two models of light neutrino masses. The first model utilizes an inverse seesaw mechanism with extra gauge singlets, while the second model assumes $R$ symmetry violation at nonrenormalizable level in the superpotential and employs the type-I seesaw mechanism. We discuss the $R$-symmetric model which leads to proton decay modes in the observable range mediated by color triplets from the Higgs multiplets. The distinctive predictions of various branching fractions and comparison with $SU(5)$ are presented. Especially, a unique prediction for  $p \rightarrow K^+ \overline{\nu}$ decay is found to serve as an additional discriminator between the present model and previous models of flipped $SU(5)$ \cite{Ellis:2020qad,Hamaguchi:2020tet,Rehman:2018gnr} where this mode is highly suppressed. An additional $Z_4$ symmetry can naturally generate  intermediate scale masses for the color triplets from the Higgs 5-plets. This is in contrast to another $R$-symmetric model recently considered in \cite{Hamaguchi:2020tet} where one of the color-triplets in Higgs 5-plet becomes naturally light and contributes only to the charged lepton channels. Lastly, in the present model a successful realization of gauge coupling unification is achieved in the presence of intermediate mass color triplets.

The layout of this paper is as follows: In Sec.~\ref{Mod} we briefly describe the flipped $SU(5)$ model including its field content, the $R$-symmetric superpotential and some of its uniquely attractive features. Two models of neutrino masses are described in Sec.~III. One is mostly based on $R$-symmetric interactions, and the second model assumes $R$-symmetry violation in the superpotential at nonrenormalizable level. In Sec.~\ref{U1R1} we discuss proton decay in $R$-symmetric flipped $SU(5)$ model mediated via both color triplets and the superheavy gauge bosons. We mainly focus on mediation by the color triplets  which occurs via the chirality nonflipping operators of  type $LLRR$ from the renormalizable interactions. 
The estimates for the proton partial lifetimes for the various channels are presented in the observable range of Hyper-K along with the lower bounds on the color triplet masses and relevant couplings. The role of an additional $Z_4$ symmetry for naturally realizing intermediate mass for the color triplets is briefly highlighted. In Sec.~\ref{U1R1SB} we discuss proton decay arising from the mixing of color-triplets in the 5-plet and 10-plet Higgs which becomes possible due to $R$-symmetry breaking effects.  This decay is especially relevant for the second model which allows $R$-symmetry breaking terms at nonrenormalizable level to generate the right handed neutrino masses.  The issue of gauge coupling unification in the presence of these intermediate mass color triplets is discussed in Sec.~VI. Our conclusions are summarized in Sec. \ref{con}.
\section{Supersymmetric Flipped $\bm{SU(5)}$ model}\label{Mod}
The Flipped $SU(5)$ gauge group is defined as $FSU(5) \equiv SU(5)\times U(1)_X$ \cite{DeRujula:1980qc,Georgi:1980pw,Barr:1981qv,Derendinger:1983aj,Antoniadis:1987dx,Barr:1988yj,Shafi:1998dv}.  The MSSM matter superfields including the right handed neutrino superfield ($N^c$) belong in the $10^1,\overline{5}^{ -3}$ and $1^5$ representations of $FSU(5)$. Here and later, if necessary, the $U(1)_X$ charge, $q(X)$, of $FSU(5)$ representations are labeled with superscripts. In contrast to $SU(5)$, the right handed neutrinos are required by the gauge symmetry in $FSU(5)$. The conjugate pair of GUT Higgs superfields, $10_H^1,\,\overline{10}_H^{-1}$ are used to break the $FSU(5)$ gauge symmetry to the minimal supersymmetric standard model (MSSM) gauge symmetry at the GUT scale $(\simeq  10^{16}\ \text{GeV})$. The electroweak breaking is accomplished through the electroweak doublets, $H_d,\, H_u$ present in $5_h^{-2},\,\overline{5}_h^{2}$.
The decomposition of various $FSU(5)$ multiplets into their MSSM  components are shown in table~\ref{assign1}. It is clear from the table that we can obtain the MSSM decomposition of $FSU(5)$ mutliplets by flipping $U^c \leftrightarrow D^c$ and $E^c\leftrightarrow N^c$ in the corresponding multiplets of the standard $SU(5)$ model \cite{Barr:1981qv}.

\begin{table}[ht]
\begin{center}
\caption{\label{assign1} The superfield content of $FSU(5)$ and charge assignments under MSSM and  $U(1)_R \times \mathbb{Z}_2$.}
\begin{tabular}{| >{\centering\arraybackslash}m{2.5cm}| | >{\centering\arraybackslash}m{4.2cm} | >{\centering\arraybackslash} m{2cm} |>{\centering\arraybackslash} m{1.5cm} | }
\hline
\vspace{0.1cm}
$SU(5)^{q(X)}$&$3_c\times 2_L \times1_Y$&$q(R)$&$\mathbb{Z}_2$\\
\hline 
\hline  
\vspace{0.1cm}
 $10^{1}$  & $Q ( 3\ \ \  2\ \ \ 1/6)$& & \\
 & $D^c ( \overline3\ \ \ 1\ \ \ 1/3)$& $0 $ &$-1$  \\
  & $N^c ( 1\ \ \ 1\ \ \ 0)$& & \\
\hline
 \vspace{0.1cm}
$ \overline5^{-3}$ & $U^c ( \overline{3}\ \ \ 1\ -2/3)$& & \\
& $L ( 1\ \ \ 2\ \ \ -1/2)$  & $0$ &$-1$ \\
\hline
 \vspace{0.1cm}
$ 1^{5}$ & $E^c ( 1\ \ \ 1\ \ \ 1)$& $0 $& $-1$ \\
\hline
\hline
\vspace{0.1cm}
 $10_H^{1}$  & $Q_H ( 3\ \ \ 2\ \ \ 1/6)$& & \\
 & $D^c_H (\overline{3}\ \ \ 1\ \ \ 1/3)$& $0 $& $+1$ \\
  & $N^c_H ( 1\ \ \ 1\ \ \ 0)$&  & \\
\hline
\vspace{0.1cm}
 $\overline{10}_H^{-1}$  & $\overline{Q_H} ( \overline{3}\ \ \ 2\ \ \ -1/6)$& & \\
 & $\overline {D^c_{H}} (3\ \ \ 1 \ \ \ -1/3)$& $0$& $+1$ \\
  & $\overline {N^c_H}\  ( 1\ \ \ 1\ \ \  0)$& & \\
\hline
\vspace{0.1cm}
 $5_h^{-2}$  & $D_h ( 3\ \ \ 1\ \ \ -1/3)$& &\\
 & $H_d ( 1\ \ \ 2\ \ \ -1/2)$&$1 $&$+1$  \\
\hline
\vspace{0.1cm}
 $\overline{5}_h^{2}$  & $\overline{D_h} ( \overline{3}\ \ \ 1\ \ \ 1/3)$& &\\
 & $ H_u (1\ \ \ 2\ \ \ 1/2)$&$1 $& $+1$  \\
\hline
\vspace{0.1cm}
$ S$ & $S(1\ \ \ 1\ \ \ 0)$ & $1$&$+1$ \\
\hline
\end{tabular}
\end{center}
\end{table}

The superpotential suitable for supersymmetric hybrid inflation in $FSU(5)$ with the additional $U(1)_R \times \mathbb{Z}_2$ symmetry listed in table~\ref{assign1} is given by \cite{Kyae:2005nv,Rehman:2009yj,Rehman:2018nsn}
\begin{align} \label{sp1}
W &=\kappa S \left(10^1_{H}\overline{10}^{-1}_{H} - M^2 \right) \nonumber \\
&+\frac{\lambda}{8} \, 10^1_{H}10^1_{H}5^{-2}_{h}+\frac{\overline{\lambda}}{8} \, \overline{10}^{-1}_{H} \overline{10}^{-1}_{H} \overline{5}^2_{h} \nonumber 
\\
& + \frac{1}{8} \, y^{(d)}_{ij}10^1_{i}10^1_{j}5^{-2}_{h}+y^{(u,\nu)}_{ij}10^1_{i}\overline{5}^{-3}_{j}\overline{5}^{2}_{h}+y^{(e)}_{ij}1^5_{i}\overline{5}^{-3}_{j}5^{-2}_{h} + W_{HN},
\end{align}
where $\lambda,\, \overline{\lambda} \text{ and } \kappa$ are real and positive dimensionless couplings. The $SU(5)$ gauge indices will be suppressed. The first term of Eq.~(\ref{sp1}) is relevant for hybrid inflation with the scalar component of the gauge singlet superfield $S$ playing the role of inflaton. The completion of hybrid inflation is followed by the $FSU(5)$ gauge symmetry breaking as the conjugate pair of 10-plet Higgses, $10^1_H, \overline{10}^{-1}_H$, acquires a nonzero vacuum expectation values (vev),  $\langle10^1_H \overline{10}^{-1}_H \rangle = \langle N_H^c \overline{N_H^c} \rangle =M^2$, along their respective MSSM singlet directions. Here the mass parameter $M$ defines the gauge symmetry breaking scale and is related to the unified gauge coupling, $g_{32}=g_5$, of $SU(3)_c\times SU(2)_L \subset SU(5)$  at the GUT scale $M_G = g_{5}\,M \approx 10^{16}$~GeV.

The  terms in the second line of Eq.~(\ref{sp1}) provide heavy masses $\lambda\,M$ and $\overline{\lambda}\,M$ respectively to the color triplet pairs ($D_H^c,\, D_h$) and ($\overline{D_H^c},\,\overline{D_h}$) residing in the Higgs superfield pairs, $(10^1_H,\,5^{-2}_h)$ and $(\overline{10}^{-1}_H,\,\overline{5}^2_h)$. It is important to note that the electroweak doublets, ($H_u,\,H_d$), in $(5^{-2}_h,\overline{5}^2_h)$ do not acquire mass from these terms. This ultimately provides the solution of doublet-triplet splitting problem via the missing partner mechanism \cite{Antoniadis:1987dx}. The significance of $U(1)_{R}$ symmetry is quite evident here as it forbids the $5^{-2}_h\,\overline{5}^2_h$ term to all orders while keeping the electroweak Higgs doublets massless, and by also avoiding dimension five proton decay mediated via the expected $5^{-2}_h \, \overline{5}^2_h$ mass term \cite{Kyae:2005nv}. The MSSM $\mu$ term, however, is assumed to be generated by the Giudice-Masiero mechanism \cite{Giudice:1988yz}. Note that the $U(1)_{R}$ symmetry also forbids the quadratic and cubic terms of $S$ for successful realization of susy hybrid inflation.

The Yukawa couplings, $y^{(u,\nu)}_{ij}, y^{(d)}_{ij}, y^{(e)}_{ij}$, in third line of Eq.~(\ref{sp1}) provide the Dirac masses for all fermions. The discussion of tiny neutrino masses and its possible connection with proton decay is included in the next sections. Some additional terms such as, $S\overline{10}^{-1}_{H}10^1_i$, $10^1_H\overline{5}^2_h\overline{5}^{-3}_i$ and $10^1_H 5^{-2}_h 10^1_i$, appear at the renormalizable level. Although $U(1)_{R}$ symmetric  these terms are forbidden by $\mathbb{Z}_{2}$ matter parity. The key feature of $\mathbb{Z}_{2}$ matter parity lies in making the lightest supersymmetric particle a dark matter candidate while forbidding the dangerous dimension four proton decay terms. The last term, $W_{HN}$, in 
Eq.~(\ref{sp1}) is responsible for generating the heavy Majorana neutrino masses necessary for the implementation of seesaw mechanism as described in the next section.
\section{Neutrino Masses}
In order to accommodate the light neutrino masses responsible for solar and atmospheric neutrino oscillations \cite{Esteban:2020cvm,Tanabashi:2018oca}, we can employ a inverse seesaw mechanism \cite{Mohapatra:1986bd,Malinsky:2005bi,Deppisch:2004fa} with the help of extra gauge singlet superfields $\mathcal{S}_a$ which have odd matter-parity with $R(\mathcal{S}_a)=1$. This allows us to include the following additional term in the superpotential at renormalizable level,
\begin{equation}
W^{I}_{HN} = \frac{\gamma_{ai}}{2} \mathcal{S}_a 10^1_i \overline{10}^1_H,
\end{equation}
where $i,a=1,2,3$. Other terms at the nonrenormalizable level relevant for proton decay are $\mathcal{S}_a 10^1_H 10^1 10^1 \bar{5}^{-3}$ and $\mathcal{S}_a 10^1_H \bar{5}^{-3} \bar{5}^{-3} 1^5$. However, their contribution to proton decay rates is highly suppressed. To implement a double seesaw mechanism we also need a mass term for the gauge singlet superfields $\mathcal{S}_a$. However, an explicit mass term, $\mu_{ab} \,\mathcal{S}_a \mathcal{S}_b$, is not allowed due to $R$-symmetry. We, therefore, include a spurion gauge singlet superfield $\Sigma$ through the K\"ahler potential term, $y_{ab} \frac{\Sigma^{\dagger}}{m_P} \mathcal{S}_a \mathcal{S}_b + h.c$, assuming even matter parity for $\Sigma$ with $R(\Sigma)=2$. An intermediate scale supersymmetry breaking by the spurion field, $\langle F_{\Sigma} \rangle \sim m_{3/2}m_P$, can lead to $\mu_{ab} \sim y_{ab} m_{3/2}$. 

A mass matrix for neutrinos and gauge singlet fields  in the $({N},\, {N}^c,\, \mathcal{S})$ basis can now be  written as 
\begin{equation}
M_{(N,\, N^c,\, \mathcal{S})} = \begin{pmatrix}
 0 & m_{(u,\nu)} & 0 \\
 m_{(u,\nu)} & 0 &
 \gamma M \\
0 & \gamma M & \mu
\end{pmatrix},
\end{equation}
where we adopt a basis in which both $m_{(u,\nu)} = y^{(u,\nu)} \upsilon_u$ and $\mu$ are real and diagonal, $m_{(u,\nu)} \simeq \text{ diag}(m_u, m_c, m_t)$
and $\mu = \text{ diag}(\mu_1, \mu_2, \mu_3)$. Applying the inverse seesaw mechanism with $\mu_a \ll | \gamma_{ja}M |$, we obtain the light neutrino mass matrix,
\begin{equation}
m_{\nu} = \frac{m^T_{(u,\nu)}  (\gamma^T)^{-1}
 \, \mu \, (\gamma)^{-1} m_{(u,\nu)}}{M^2} ,
\label{eq:mnu}
\end{equation}
which is diagonalized by a unitary matrix $U_N$, namely $m_{\nu}^{diag} = U_N^* m_{\nu}  U_N^\dagger$. This is in contrast to the double seesaw mechanism where $\mu_a \gg | \gamma_{ja}M |$ is assumed. See Refs.~\cite{Ellis:2019jha,Ellis:2019opr} for a recent analogous treatment of double seesaw mechanism in an inflation model based on $FSU(5)$ gauge symmetry. To provide an estimate for the relevant couplings, we set $M \simeq 1.4 \times 10^{16}$~GeV and $\mu_a \sim 10$~TeV and obtain normal mass hierarchy for the light neutrinos, $m^{diag}_{\nu} \simeq \text{ diag}(1.1 \times 10^{-7} , 0.0086,  0.05)$~eV, for $\gamma \simeq \text{ diag}(1.6 \times 10^{-9} , 2.9 \times 10^{-9}, 1.8 \times 10^{-7})$. As we discuss in the next section, with the help of an additional $Z_4$ symmetry these values of $\gamma$ couplings can be boosted by the factor $(m_P/M)^2\sim 10^4$.
The mixed states of $N^c$ and $\mathcal{S}_a$ obtain masses of order, $\gamma M \simeq \text{ diag}(2.3 \times 10^{7} , 4.1 \times 10^{7}, 2.5 \times 10^{9})$~GeV.

In general for a given matrix $\gamma_{ia}$, the mixing matrix $U_N$ can be determined as a function of $\mu_a$. However, for numerical estimates we will assume normal-ordered (NO) light neutrino masses with $U_N$ equal to a unit matrix. This also allows us to write the mixing matrix $U_L = U_{\rm PMNS}^* U_N^* \simeq U_{\rm PMNS}^*$ in term of the Pontecorvo-Maki-Nakagawa-Sakata (PMNS) matrix, $U_{\rm PMNS} = U_L^{*} U_N^{\dagger}$. This enables us to estimate all relevant proton decay rates mediated by the color triplets in the Higgs 5-plets as discussed in the next section.

An alternative interesting possibility for generating light neutrino masses can be realized by allowing explicit $U(1)_{R}$ symmetry breaking terms at the nonrenormalizable level \cite{Civiletti:2013cra}. As $U(1)_R$ is a global symmetry it could be broken in the hidden sector while mediating breaking effects to the visible sector via gravitational interactions. We will assume that the $R$-symmetry breaking occurs in such a way that it only allows terms with $R=0$ charge in the superpotential at the nonrenormalizable level. With $\mathbb{Z}_2$ matter parity present only an even number of matter superfields  appear with the 10-plet Higgs fields. To leading order, the following terms are allowed in the superpotential,
\begin{eqnarray}
W^{II}_{HN} &\supset & 
\frac{\gamma_0}{4} \left(\frac{(10^1_H \overline{10}^{-1}_H)^2}{m_P}\right) + \frac{\gamma_1}{4} \left(\frac{ {(10^1 \overline{10}^{-1}_H})^2}{m_P}\right) + \frac{\gamma_2}{4} \left(\frac{ {(10^1 \overline{10}^{-1}_H}) \cdot ( 10^1 \overline{10}^{-1}_H)}{m_P}\right)  \nonumber \\
&+&    \frac{\gamma_3}{8} \left(\frac{10^1_H 10^1_H 10^1 \overline{5}^{-3}}{m_P}\right) +\frac{\gamma_4}{8} \left(\frac{{\overline{10}^{-1}_H}\overline{10}^{-1}_H\ \overline{5}^{-3} 1^5}{m_P}\right),\label{WII}
\end{eqnarray}
where $\gamma_k$, with  $k=0,1,2,3,4$, are the dimensionless matrices with family indices suppressed. This model is briefly discussed in Sec.~V where rapid proton decay mediated by the color triplets in the Higgs 5-plets and 10-plets strongly restricts the couplings, $\gamma_{2,3,4} \lesssim  10^{-5}$.

A neutrino mass matrix in the $(N,\, N^c)$ basis can be written as 
\begin{equation}
M_{(N,\, N^c)} = \begin{pmatrix}
 0 & m_{(u,\nu)}  \\
 m_{(u,\nu)} & M_{\nu^c} 
\end{pmatrix},
\end{equation}
where the third term in the superpotential $W^{II}_{HN}$  provides the mass matrix, $M_{\nu^c} =  \frac{\gamma_1 M^2}{m_P}$, for the right handed neutrinos assuming $\gamma_2 \ll \gamma_1$.
The light neutrino mass matrix is obtained via the standard seesaw mechanism \cite{Minkowski:1977sc},
\begin{equation}
m_{\nu} = m^T_{(u,\nu)}  (M_{\nu^c})^{-1}
 \, m_{(u,\nu)} ,
\label{eq:mnu2}
\end{equation}
and is diagonalized by a unitary matrix $U_N$ as $m_{\nu}^{diag} = U_N^* m_{\nu}  U_N^\dagger$. For numerical estimates in this second model of neutrino masses we adopt the basis where $(\gamma_1)_{ij}$ is real and diagonal, $(\gamma_1)_{ij}=(\gamma_1)_{i}\delta_{ij}$, with $(M_{\nu^c})_{ij} = M_{i}\delta_{ij}$ and $M_{i} = (\gamma_1)_{i} M^2/m_P$. For a normal mass hierarchy of the light neutrinos, $(m^{diag}_{\nu}) \simeq \text{ diag}(1.1 \times 10^{-7} ,  0.0086, 0.05)$~eV, and with $M \simeq 1.4 \times 10^{16}$~GeV, we obtain $M_{\nu^c} \simeq \text{ diag}(5.1 \times 10^{10}, 1.7 \times 10^{11}, 6.0 \times 10^{14})$~GeV for $\gamma_1\simeq \text{ diag}(6.3 \times 10^{-4}, 2.0 \times 10^{-3}, 7.4)$. These values of the  right handed neutrino masses are significantly larger than the corresponding estimate of heavy neutrino masses obtained in the inverse seesaw mechanism described above. This scenario can be naturally incorporated in hybrid inflation models with successful reheating and nonthermal leptogenesis \cite{Senoguz:2003hc}.
\section{proton decay in $\bm{FSU(5)}$ with $\bm{U(1)_{R}}$ symmetry}\label{U1R1}
Proton decay in $FSU(5)$ mediated by the superheavy gauge bosons has been extensively studied in the past \cite{Barr:1982pk,Ellis:1993ks,Ellis:1995at,Ellis:2002vk,Dorsner:2004xx,Li:2010ar} mostly in comparison with the unflipped $SU(5)$ model. In a recent paper \cite{Ellis:2020qad} this is discussed in a no-scale supersymmetric $FSU(5)$  inflation model with an approximate $Z_2$ symmetry and modified $R$ parity. In this section we will explore proton decay in an $R$-symmetric $FSU(5)$ model suitable for susy hybrid inflation model. As emphasized earlier the $U(1)_R \times \mathbb{Z}_2$ symmetry plays an important role in suppressing various operators that mediate rapid proton decay. For example, the dimension four rapid proton decay mediated through the color triplet, $D^c\subset 10^1$, can appear at nonrenormalizable level via the following operators,  
\begin{eqnarray}
\frac{S 10_H 10_i 10_j \overline{5}_k}{m_P^2}  &\supset& \frac{\langle S \rangle M}{m_P^2}
 \left(Q_i D^c_j L_k + D^c_i D^c_j U^c_k \right), \\ 
\frac{S 10_H \overline{5}_i \overline{5}_j \overline{1}_k }{m_P^2}  &\supset& \frac{\langle S \rangle M}{m_P^2} \left( L_i L_j E^c_k \right).
\end{eqnarray}
Without the $S$ field and with no $U(1)_R$ symmetry these operators can lead to fast proton decay incompatible with the experimental observations. The presence of $S$ is required by the $U(1)_R$ symmetry which makes these operators highly suppressed as the $S$ field is expected to acquire  a vev of order TeV scale from the soft susy breaking terms \cite{Dvali:1997uq}. Note that these operators are also forbidden by the $\mathbb{Z}_2$ matter parity even if we allow $R$-symmetry breaking operators at nonrenormalizable level as discussed in the previous section. The GUT scale mass terms for Higgs 5-plets, $5_h\overline{5}_h$, and Higgs 10-plets, $10_H\overline{10}_H$, are also not allowed due to $U(1)_R$ symmetry and which may otherwise mediate dimension five rapid proton decay.

For proton decay via dimension five and dimension six  operators we mainly focus on the mediation by color triplets in the conjugate pairs of Higgs superfields, $(5_h^{-2}, \overline{5}_h^2)$ and $(10^1_H, \overline{10}^{-1}_H)$. In general, these color triplets can contribute to proton decay via operators of chirality types $LLLL$, $RRRR$  and  $LLRR$, as discussed in a recent paper on 4-2-2 model \cite{Lazarides:2020bgy}. In our model $R$ symmetry with renormalizable interactions only allows the chirality nonflipping modes which reduce to the following four Fermi operators of $LLRR$ chirality generated via color triplet exchange from $5_h^{-2}, \, \overline{5}_h^2$,
\begin{eqnarray}
10^1\,\overline{5}^{-3} \, {10^{1}}^{\dagger} \, {\overline{5}^{-3}}^{\dagger} & \supset& Q \, L \, {U^c}^{\dagger} \, {D^c}^{\dagger} = (U \, E + D \, N )\, {U^c}^{\dagger} \, {D^c}^{\dagger}, \label{fermi1}\\
10^1 \, 10^1 \, {\overline{5}^{-3}}^{\dagger} \, {1^{5}}^{\dagger} & \supset&  Q \, Q \, {U^c}^{\dagger} \, {E^c}^{\dagger} = (U \, D + D \, U) \, {U^c}^{\dagger} \, {E^c}^{\dagger}. \label{fermi2}
\end{eqnarray} 
Later we also discuss the proton decay mediation by the color triplets from $10_H, \, \overline{10_H}$ by allowing explicit $R$-symmetry breaking terms with $R$-charge zero at the nonrenormalizable level.

\begin{figure}[ht]\centering
\subfloat[]{\includegraphics[width=1.87in]{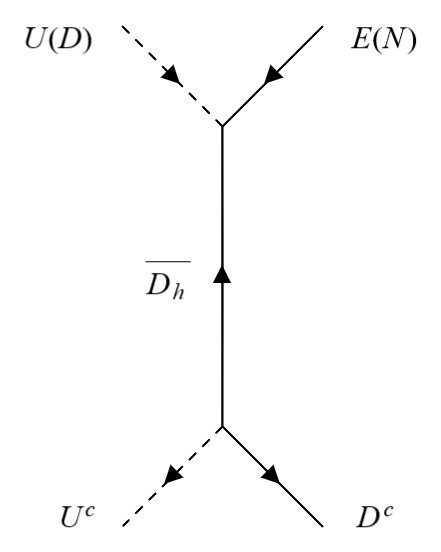}\label{1a}}\hspace{2.0cm} 
\subfloat[]{\includegraphics[width=1.62in]{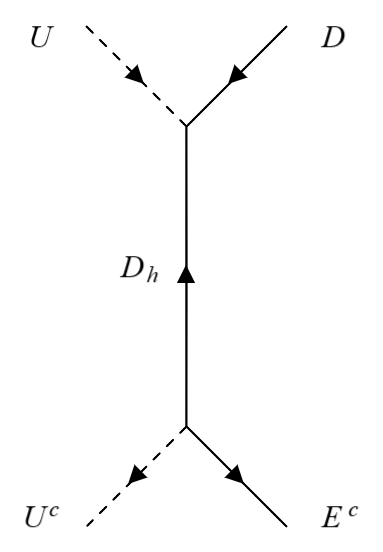}\label{2a}} \\
\subfloat[]{\includegraphics[width=1.77in]{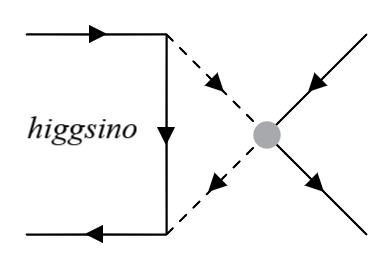}\label{3a}}\hspace{2.0cm} 
\subfloat[]{\includegraphics[width=1.77in]{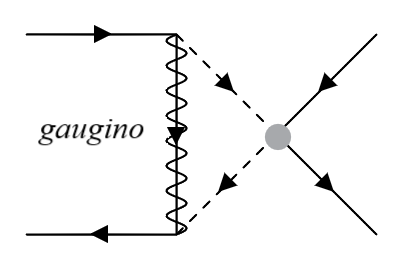}\label{4a}}
\caption{Dimension five proton decay diagrams corresponding to effective operators $10\,\overline{5}\,{10}^{\dagger}\,{\overline{5}}^{\dagger}$ and $10\,10\, {\overline{5}}^{\dagger}\,{1}^{\dagger}$ mediated by fermionic color triplets $\overline{D_h} \subset \overline{5}_h$ and $D_h \subset {5_h}$ are shown in panels (a) and (b) respectively. The dashed and solid lines represent bosons and fermions respectively. The bosonic and fermionic character of external lines can be interchanged at each vertex in the first two diagrams. The generic loop diagrams for dimension five proton decay with chirality nonflipping mediation by higgsino or gaugino are shown in the last two panels (c) and (d) respectively. Here the dot represents the effective dimension five operator once the heavy color triplet mediating fields have been integrated out.}
\label{fig:dim511}
\end{figure}

\begin{figure}[t]\centering
	\subfloat[]{\includegraphics[width=1.72in]{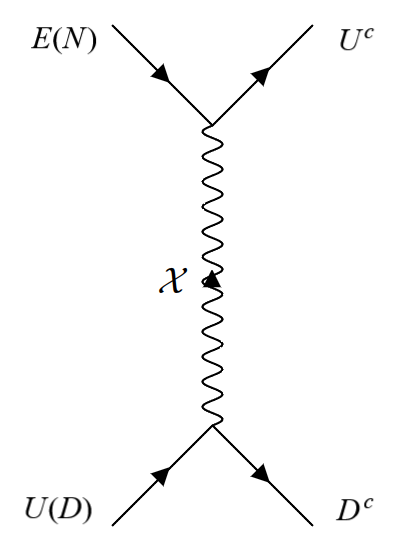}\label{1b}}\qquad
	\subfloat[]{\includegraphics[width=1.89in]{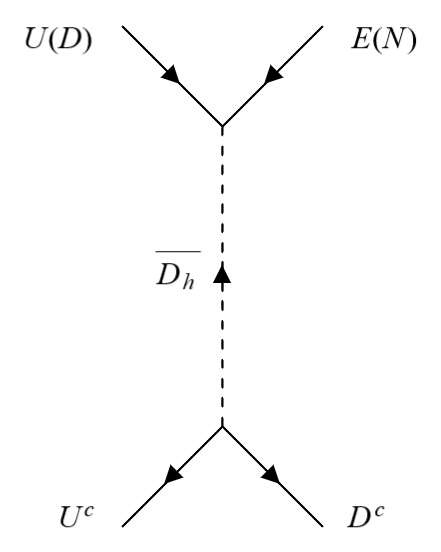}\label{2b}}\qquad
	\subfloat[]{\includegraphics[width=1.65in]{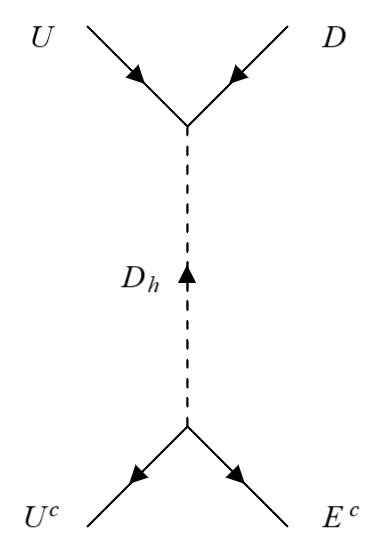}\label{3b}}
	\caption{Dimension six proton decay diagrams corresponding to effective operators $10\,\overline{5}\,{10}^{\dagger}\,{\overline{5}}^{\dagger}$ and $10\,10\, {\overline{5}}^{\dagger}\,{1}^{\dagger}$ mediated by gauge bosons and scalar color triplets $\overline{D_h} \subset \overline{5}_h$ and $D_h \subset {5_h}$. The wavy, dashed and solid lines represent vector bosons, scalars and fermions respectively.}
	\label{fig:dim611}
\end{figure}

The Feynman diagrams for dimension five and dimension six proton decay operators are shown in Figs.~(\ref{fig:dim511}) and (\ref{fig:dim611}).  The dashed and solid lines represent bosons and fermions respectively. In dimension five diagrams the bosonic and fermionic character of external lines can be interchanged at each vertex.
The external dashed lines in the dimension five diagrams form a loop with chirality nonflipping higgsino or gaugino mediation as shown in Figs.~(\ref{3a}) and (\ref{4a}). Here the dot represents the effective dimension five operator once the heavy color triplet mediating fields has been integrated out. Below supersymmetry breaking scale, $M_{SUSY}$, we finally obtain the four Fermi effective operators of the kind $LLRR$ already expressed in Eqs.~(\ref{fermi1}) and (\ref{fermi2}). 

A word of caution is in order regarding the proper use of the terminology for the operators of type $LLRR$ involving two scalars and two fermions. These are actually dimension six operators which can be seen from the following K\"ahler potential term,
\begin{equation}
{\cal L} \supset \frac{1}{m_P^2} \int d^4\theta \, \Phi \Phi^{\dagger} \Phi \Phi^{\dagger} \supset 
\frac{1}{m_P^2} \overline{\psi} \, \slashed{\partial} \, \psi  \, \phi^* \phi
\end{equation}
where $\Phi \supset \phi(x)+\sqrt{2}\theta\psi(x)-i\frac{1}{\sqrt{2}}\theta^2\partial^{\mu}\psi(x)\sigma_{\mu}\bar{\theta}$. Therefore, the corresponding effective operator with two scalars and two fermions always contains a derivative. It is also clear that the chirality non-flipping fermion propagator picks up the momentum of the fermion for the diagram given in Figs.~(\ref{1a},\,\ref{2a}). Therefore, the dimension of chirality non-flipping  operators of type $LLRR$ with two scalars and two fermions is six.

The Yukawa terms in the superpotential $W$ (Eq.~(\ref{sp1})) relevant for proton decay mediated by the color triplets $(D_h,\,\overline{D_h})$ of $(5_h^{-2},\overline{5}_h^2)$ can be expressed in terms of mass eigenstates as
\begin{eqnarray}
W &\supset & L \left( U_L \, y_D^{(u,\nu)} \right) Q\, \overline{D_h} + U^c \left( y_D^{(u,\nu)}V P^* \right) D^c \overline{D_h} \\
&-& \frac{1}{2}\,Q \left( V^* P \, y_D^{(d)} \, V^{\dagger} \right) Q \, D_h  +  U^c \left(U_L^{\dagger} \, y_D^{(e)} \right) E^c D_h,\label{req2}
\end{eqnarray}
with the diagonal Yukawa couplings, $y_D^{(u,\nu)},\,y_D^{(d)},\,y_D^{(e)}$, given by
\begin{equation}
y^{(u,\nu)} = y_D^{(u,\nu)},\,  y^{(d)} = V^* \,P \, y_D^{(d)} \, \ V^{\dagger} ,\,  y^{(e)} = U_L^{\dagger} \, y_D^{(e)} \, U_{E^c}^{\dagger}.
\end{equation}
The $FSU(5)$ supermultiplets are expressed in terms of the following mass eigenstates \cite{Ellis:1993ks,Ellis:2020qad}
\begin{align}
 10^1 & \ni \left\{ Q , \, V P^* D^c, \, U_{N^c}
 N^c \right\} \text{ with } Q = \begin{pmatrix}
 U \\ V D
\end{pmatrix}, \nonumber \\
 \bar{5}^{-3} & \ni \left\{
U^c, \, U_L^T L \right\}  \text{ with } L = \begin{pmatrix}
 U_{\rm PMNS} N \\ E
\end{pmatrix} \text{ and } U_{\rm PMNS} = U_L^{*} U_N^{\dagger}, \nonumber \\
 1^5 &=  U_{E^c} E^c,
\label{eq:embedding}
\end{align}
where $V$ is the Cabibbo-Kobayashi-Maskawa (CKM) matrix and $P = \text{diag}(e^{i\varphi_1},e^{i\varphi_2},e^{i\varphi_3} )$ is the phase factor matrix with the condition $\sum_{i} \varphi_i = 0$~\cite{Ellis:2020qad}. 

As the amplitude of dimension five diagrams involves loop factors their contribution is generally expected to be suppressed as compared to dimension six diagrams. Therefore, we will include the contribution of color triplets only from dimension six diagrams which are generated from a combination of the Yukawa terms in the Lagrangian $\int d^2 \theta W$ and their Hermitian conjugates. Similarly, the gauge boson exchange diagram (\ref{1b}) is generated  from the following part of the K\"ahler potential \cite{Ellis:2020qad},
\begin{equation}
K \supset  \sqrt{2} \, g_5 \, \bigl(
 -(U^c)^{\dagger} \mathcal{X} (U_{L}^T L)
 +(Q)^{\dagger} \mathcal{X} ( VP^* {D}^c)
 +{\rm h.c.}
 \bigr),
\end{equation}
where $\mathcal{X}$ is the $SU(5)$ gauge vector superfield. The combined effects of the superheavy $SU(5)$ gauge boson and  color triplet mediation below their mass scales are described by the dimension six effective operators,
\begin{eqnarray}
 {\cal L}_{6}^{\rm eff}
&=& C_{6(1)}^{ijkl} \,
\bigl(U^{c}\bigr)_i^{\dagger}\,
\bigl(D^{c}\bigr)^{\dagger}_j
\, Q_k \, L_l
+ C_{6(2)}^{ijkl} \, Q_i \, Q_j \, \bigl(U^{c} \bigr)^{\dagger}_k \,
\bigl(E^{c}\bigr)^{\dagger}_l,  \\
&\supset& \bigl(U^{c}\bigr)_i^{\dagger}\,
\bigl(D^{c}\bigr)^{\dagger}_j
\, C_{6(1)}^{ijkl} ( U_k E_l + (VD)_k  (U_{PMNS}N)_l)  \nonumber \\
&+& (U_i \, (V\,D)_j + (V\,D)_i \, U_j) \, C_{6(2)}^{ijkl} \, \bigl(U^{c} \bigr)^{\dagger}_k \, \bigl(E^{c}\bigr)^{\dagger}_l,
\label{eq:l6eff}
\end{eqnarray}
where the Wilson coefficients $C^{ijkl}_{6(1,2)}$ are  given by
\begin{align}
 C^{ijkl}_{6(1)}&= e^{i\varphi_j} \left(\frac{(U_L)_{li} V_{kj}^*}{M^2}
 + \frac{(V^{\dagger} \, y_D^{(u,\nu)})_{ji}(U_L  \, y_D^{(u,\nu)})_{lk}}{M^2_{\bar{\lambda}}}\right),  \label{wilson1} \\
 C^{ijkl}_{6(2)}&=  - \left(\frac{(V^* \, P\, y_D^{(d)} \, V^{\dagger})_{ij}(U^T_L \, y_D^{(e)})_{kl}}{2 M^2_{\lambda}}\right).
\label{wilson2}
\end{align}
Here the color triplet masses are written as $M_{\lambda}=\lambda\ M$ and $ M_{\bar{\lambda}}=\overline{\lambda} \ M$. The first term in $ C^{ijkl}_{6(1)}$ is the contribution from the gauge boson exchange diagram (\ref{1b}) which has been studied recently in \cite{Ellis:2020qad} for an inflation based model. The contribution of the first term in $ C^{ijkl}_{6(2)}$ arises from the $D_h$ color triplet exchange diagram (\ref{2b}). This contribution has been studied more recently in an $R$-symmetric flipped $SU(5)$ model \cite{Hamaguchi:2020tet} which naturally predicts $M_{\lambda}$ to be of intermediate scale. With $M_{\bar{\lambda}}$ of order $M_G$ only the charged lepton channels are predicted to lie in the observable range of future experiments at Hyper-K \cite{Abe:2018uyc}.  The contribution of the second term in $ C^{ijkl}_{6(1)}$ arises from the $\overline{D_h}$ color triplet exchange diagram (\ref{3b}) and is crucial for  making a nonvanishing prediction for the $K^+\bar{\nu}$ decay channel which is usually assumed to be suppressed. The present model with an additional  $Z_4$ symmetry, as described in the next section, naturally predicts both $M_{\lambda}$ and $M_{\bar{\lambda}}$ to be of intermediate scale. This leads to distinctive proton lifetime predictions especially for the neutral lepton decay channels as described below.

The Wilson coefficients $C^{ijkl}_{6(n)}$ $(n = 1,2)$ in Eqs.~(\ref{wilson1}) and (\ref{wilson2}) are run down to low energy scales using the Renormalization Group Equations (RGEs) given in \cite{Hisano:2013ege}. The effect of one-loop RGE between the GUT scale $M_G$ and the electroweak scale $M_Z$ are encoded in the renormalization factors, $A_{S_n}$, \cite{Munoz:1986kq, Abbott:1980zj}:
\begin{eqnarray}
 A_{S_1} &=&  \prod_{i=1}^{3} \biggl[ \frac{\alpha_i({M_T})}{\alpha_i(M_{G})} \biggr]^{\frac{c^{(1)}_i}{2 b^{(3)}_i}}\times 
 \biggl[\frac{\alpha_i(M_{\text{SUSY}})}{\alpha_i({M_T})}\biggr]^{\frac{c^{(1)}_i}{2 b^{(2)}_i}}\times
\biggl[\frac{\alpha_i(M_Z)}{\alpha_i(M_{\rm SUSY})}]\biggr]^{\frac{c^{(1)^{\prime}}_i}{2 b^{(1)}_i}}, \\
 A_{S_2} &=& \prod_{i=1}^{3} \biggl[ \frac{\alpha_i({M_T})}{\alpha_i(M_{G})} \biggr]^{\frac{c^{(2)}_i}{2 b^{(3)}_i}}\times 
 \biggl[\frac{\alpha_i(M_{\text{SUSY}})}{\alpha_i({M_T})}\biggr]^{\frac{c^{(2)}_i}{2 b^{(2)}_i}}\times
\biggl[\frac{\alpha_i(M_Z)}{\alpha_i(M_{\rm SUSY})}]\biggr]^{\frac{c^{(2)^{\prime}}_i}{2 b^{(1)}_i}},
\end{eqnarray}
where $c^{(1,2)}_i$ ($c^{(1,2)^{\prime}}_i$) are the coefficients of one-loop RGEs for Wilson coefficients $C_{6(1,2)}^{ijkl}$ above (below) the SUSY scale, $M_{\rm SUSY}$, and are given as
\begin{eqnarray}
c^{(1)}_{i}&=&(-\frac{11}{15} ,\ -3,\ -\frac{8}{3} ),\,c^{(2)}_{i}=( -\frac{23}{15},\ -3 ,\ -\frac{8}{3}),  \\
c^{(1)^{\prime}}_{i}&=&(-\frac{11}{10} ,\ -\frac{9}{2},\ -4 ),\, c^{(2)^{\prime}}_{i}=( -\frac{23}{10},\ -\frac{9}{2} ,\ -4 )  .
\end{eqnarray}
The one-loop beta coefficients, $b^{(3)}_i$, $b^{(2)}_i$ and $b^{(1)}_i$, of the gauge couplings $\alpha_i=g_i^2/(4\pi)^2$ are given by
\begin{eqnarray}
b^{(3)}_{i}=b^{(2)}_{i}+n_T\,( \frac{2}{5}, \, 0,\, 1)+n_D\,( \frac{3}{5},\,1 ,\,0),\,
 b^{(2)}_{i}=(\frac{33}{5} ,\, 1,\,-3),\, b^{(1)}_{i}=(\frac{41}{10} ,-\frac{19}{6}  ,\ -7),
\end{eqnarray}
for MSSM plus light $n_T$ triplets and $n_D$ doublets, MSSM and SM content respectively. The additional $n_D$ electroweak doublets are required to achieve MSSM gauge coupling unification as discussed in the next section. For simplicity, we take the same numbers of color triplets and electroweak doublets ($n_T=n_D$) with equal mass $M_T$. The triplet mass $M_T= M_{\lambda}$ or $M_{\bar{\lambda}}$ depending upon the color triplet that makes the dominant contribution to proton decay. The renormalization factors are suplemented by another factor $A_L=1.247$ \cite{Nihei:1994tx} which is the perturbative QCD renormalization factor below the electroweak scale represented by the $Z$ boson mass, $M_Z$.

The dimension six proton decay is mediated by vector gauge bosons and the color triplets (from $5^{-2}_h$ and $\overline{5}^2_{h}$). Therefore, the decay rates for charged-lepton channels with $l^+_i=(e^+,\,\mu^+)$ become,
\begin{eqnarray}
\Gamma_{p \rightarrow \pi^0 l^+_i}& = & k_{\pi}|C_{\pi^0 l^+_i}|^2\left(A^2_{S_1}\left|\frac{1}{M^2}+\left(\frac{m_u}{\upsilon_u}\right)^2\frac{1}{M^2_{\bar{\lambda}}}\right|^2 
 +  A^2_{S_2}\left| \frac{m_{d}}{\upsilon_d} \frac{m_{l_i}}{\upsilon_d} \frac{1}{M^2_{\lambda}} \right|^2\right), \label{dr1}\\
\Gamma_{p \rightarrow K^0 l^+_i}& = & k_{K}|C_{K^0 l^+_i}|^2 \left( A^2_{S_1}\left|\frac{1}{M^2}+\left(\frac{m_u}{\upsilon_u}\right)^2\frac{1}{M^2_{\bar{\lambda}}}\right|^2 + A^2_{S_2}  \left| \frac{m_{s}}{\upsilon_d} \frac{m_{l_i}}{\upsilon_d} \frac{1}{M^2_{\lambda}} \right|^2 \right), \label{dr2}
\end{eqnarray}
where $m_p,\ m_{\pi}, \ m_{K}$ and $m_{l_i}=(m_{e},\,m_{\mu})$ are the masses of proton, pion, kaon and charged leptons $l_i$ respectively. The MSSM parameters are  $\upsilon_u=\upsilon \sin{\beta}$ and $ \upsilon_d=\upsilon \cos{\beta}$ with electroweak vev, $\upsilon=174$~GeV. Finally, the $k$- and the $C$-factors are respectively defined as
\begin{eqnarray}
k_{\pi} = \frac{m_p A_L^2}{32\pi}\left(1-\frac{m^2_{\pi}}{m^2_p}\right)^2,\quad  
k_{K}= \frac{m_p A_L^2}{32\pi}\left(1-\frac{m^2_{K}}{m^2_p}\right)^2,
\end{eqnarray}
and
\begin{equation}
C_{\pi^0 l_i} = T_{\pi^0 l_i} (U_L)_{i1} V^*_{ud},\qquad  C_{K^0 l_i} = T_{K^0 l_i} (U_L)_{i1} V^*_{us}.
\end{equation}
For convenience, the recently updated values of hadronic matrix elements $T_{ml}$ from lattice computation  \cite{Aoki:2017puj} and the corresponding Super-K bounds \cite{Tanabashi:2018oca,Miura:2016krn,Abe:2013lua,Takhistov:2016eqm,Regis:2012sn,Abe:2014mwa,Kobayashi:2005pe}, Hyper-K \cite{Abe:2018uyc} and DUNE \cite{An:2015jdp} sensitivities  are given in Table~\ref{matrix-el}.

\begin{table}[t]
\begin{center}
\caption{The Super-K bounds, Hyper-K and DUNE sensitivities and values of relevant matrix elements for various proton decay channels.}\label{matrix-el}
\scalebox{1.0}{
\begin{tabular}{| >{\centering\arraybackslash}m{1.35cm}|  >{\centering\arraybackslash}m{5.1cm} | >{\centering\arraybackslash} m{2.6cm} |>{\centering\arraybackslash} m{2.1cm} |>{\centering\arraybackslash} m{2.25cm} |>{\centering\arraybackslash} m{1.9cm} | }
\hline
\vspace{0.1cm}
 \multirow{3}{4em}{Decay channel} &\multicolumn{2}{c|}{} & Super-K  & \multicolumn{2}{c|}{ Sensitivities }\\
 & \multicolumn{2}{c|}{$T_{ml} $= Matrix element ($ \text{GeV}^2 $) }  &bound \cite{Tanabashi:2018oca}& \multicolumn{2}{c|}{($10^{34}$ years)}\\
 \cline{5-6}
 &\multicolumn{2}{c|}{}&($10^{34}$ years) & Hyper-K\cite{Abe:2018uyc}  &DUNE\cite{Acciarri:2015uup} \\
\hline 
\hline  
\vspace{0.1cm}
$e^+ \,\pi^0$ &$T_{\pi^0 e^+}=\langle \pi^0|(ud)_R u_L|p\rangle_{e^+}$& $-0.131(4)(13)$&$1.6$ &$7.8$&$--$\\
$\mu^+ \, \pi^0$&$T_{\pi^0 \mu^+}=\langle \pi^0|(ud)_R u_L|p\rangle_{\mu^+}$&$ -0.118(3)(12)$&$0.77$ &$7.7$&$--$\\
$\bar{\nu} \, K^+ $&$T^{\prime}_{(K^+ \bar{\nu})}=\langle K^+|(ud)_R s_L|p\rangle$& $-0.134(4)(14)$&$0.59$&$3.2$&$6.2$\\
&$T^{\prime \prime}_{(K^+ \bar{\nu})}=\langle K^+|(us)_R d_L|p\rangle$&$-0.049(2)(5)$&&&\\
$\bar{\nu} \, \pi^+ $&$T_{\pi^+ \bar{\nu}}=\langle \pi^+|(ud)_R d_L|p\rangle$& $-0.186(6)(18)$&$0.039$&$--$&$--$\\
$e^+ \, K^0$&$T_{K^0 e^+}=\langle K^0|(us)_R u_L|p\rangle_{e^+}$& $\ 0.103(3)(11)$& $0.1$ &$--$&$--$\\
$\mu^+ \, K^0$&$T_{K^0 \mu^+}=\langle K^0|(us)_R u_L|p\rangle_{\mu^+}$& $\ 0.099(2)(10)$&$0.16$&$--$&$--$\\
\hline
\end{tabular}}
\end{center}
\end{table}

\begin{figure}[t]\centering
\subfloat[$p \rightarrow e^+ \pi^0 $]{\includegraphics[width=3.268in]{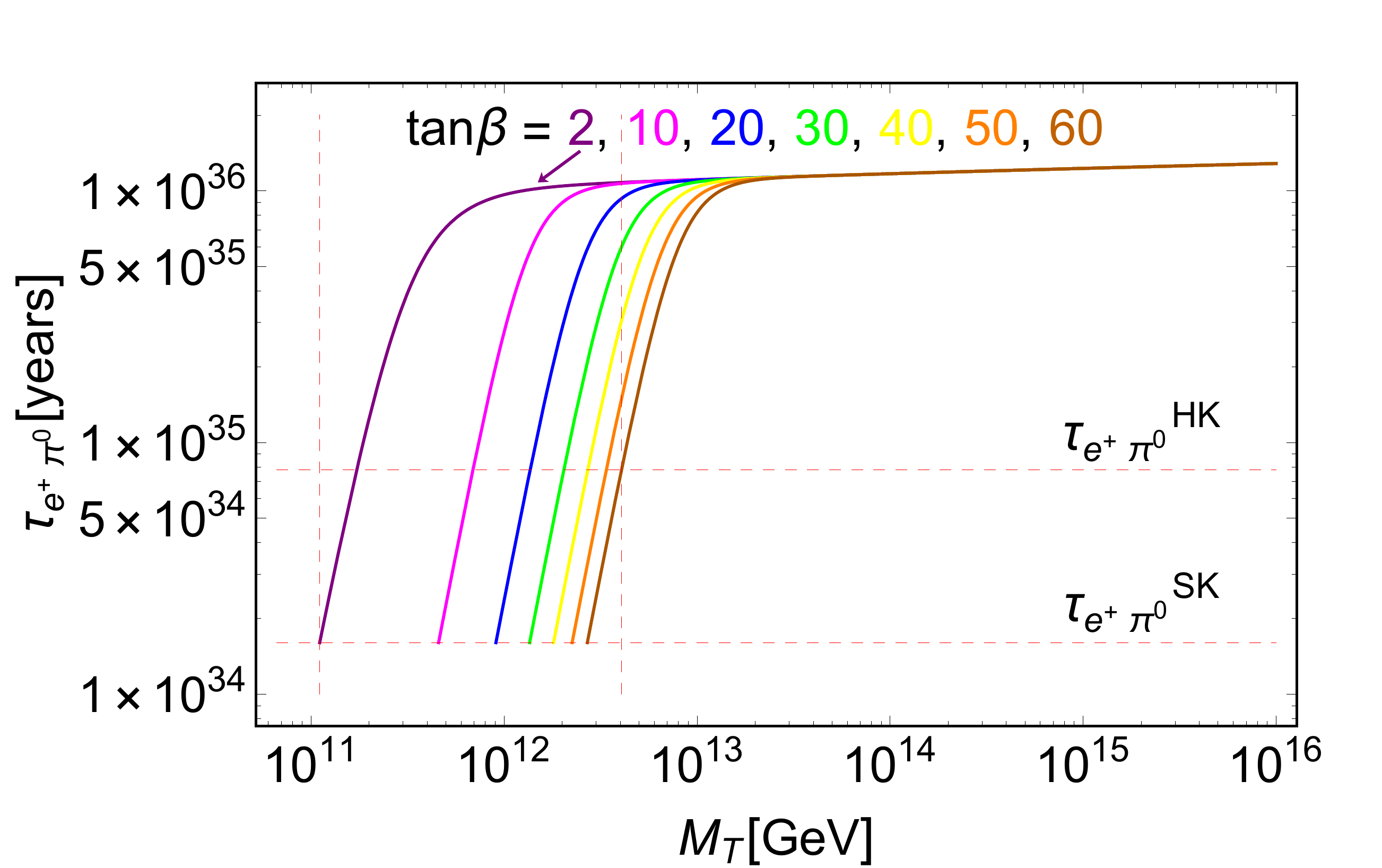}\label{epi3}} \hspace{0.1cm}
\subfloat[$p \rightarrow \mu^+ \pi^0 $]{\includegraphics[width=3.10in]{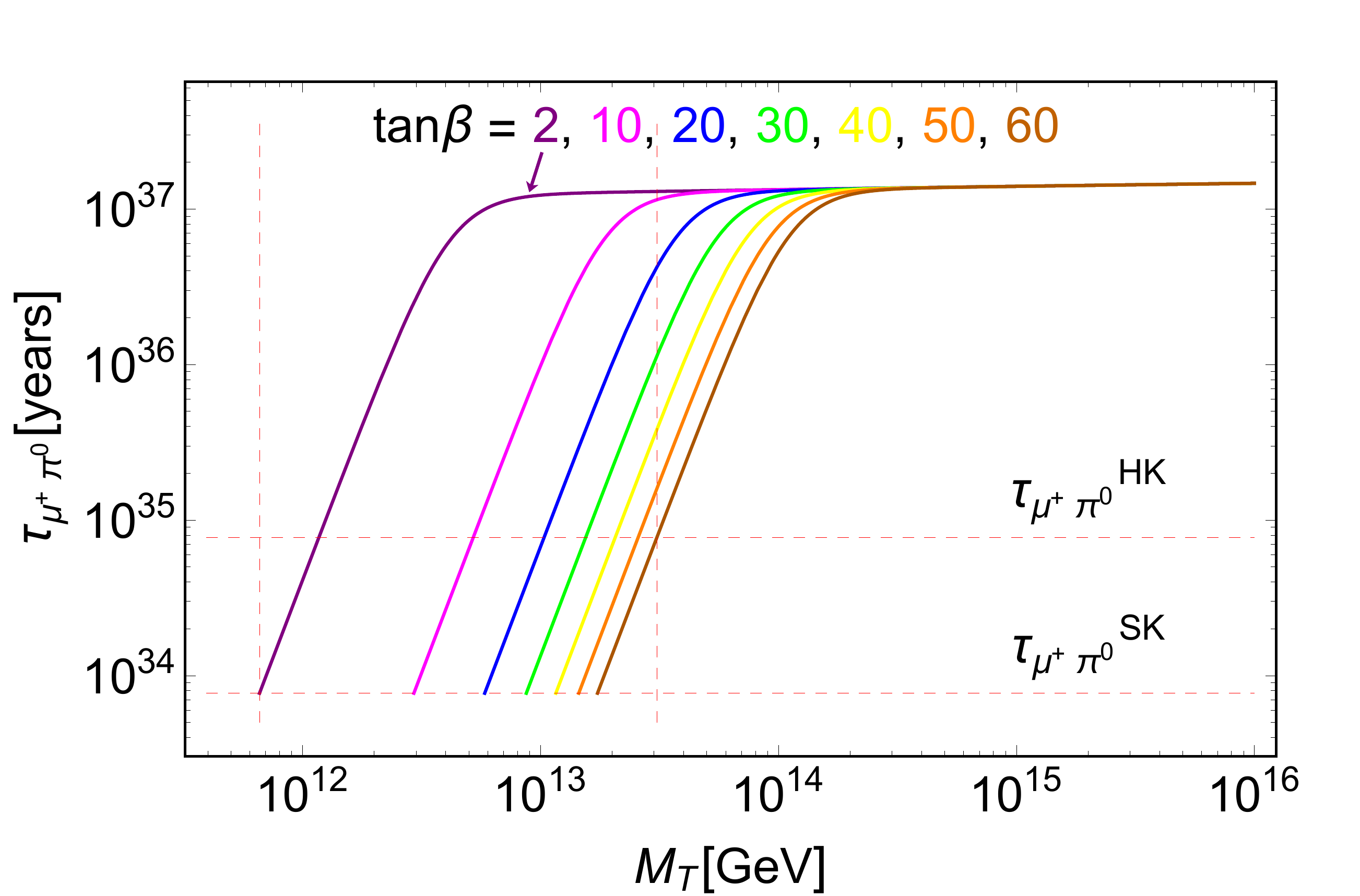}\label{mupi3}}\\
\subfloat[$p \rightarrow e^+ K^0 $]{\includegraphics[width=3.1in]{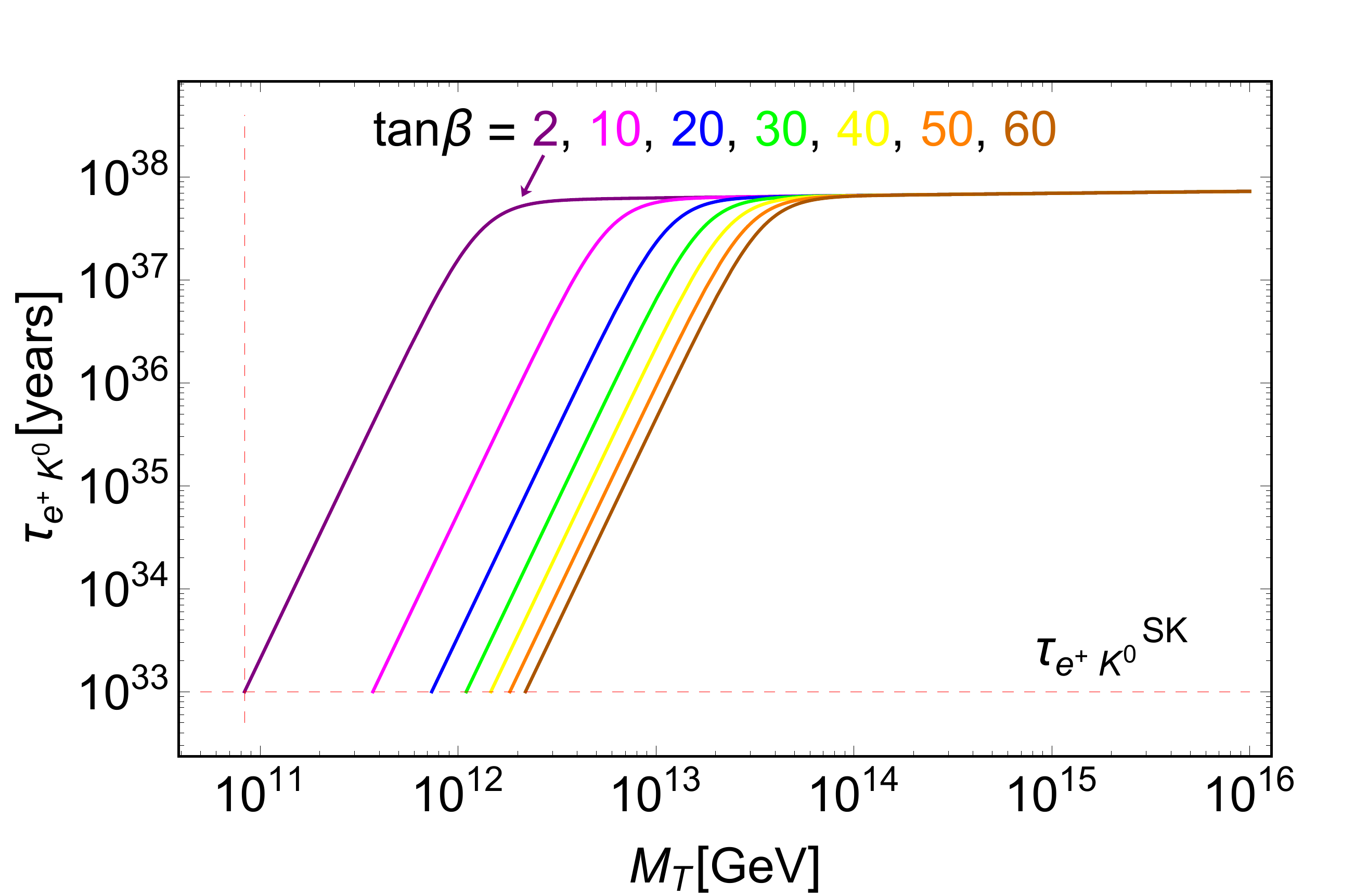}\label{epi4}} \hspace{0.1cm}
\subfloat[$p \rightarrow \mu^+ K^0 $]{\includegraphics[width=3.1in]{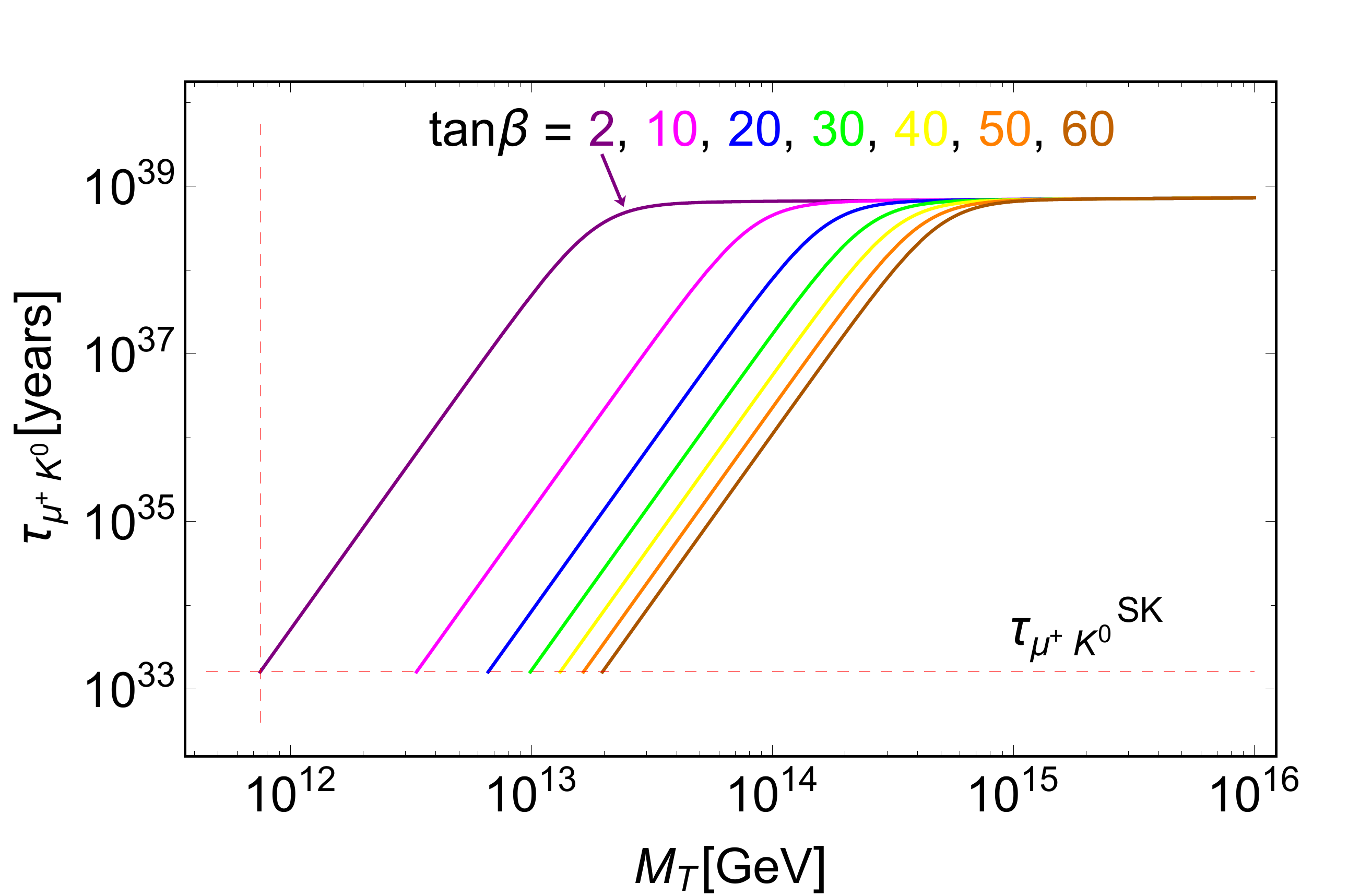}\label{mupi4}} 
\caption{The partial-lifetime estimates of proton for charged-lepton decay channels as a function of 5-plet triplet mass $M_T = M_{\bar{\lambda}} = M_{{\lambda}}$ with $\tan \beta$ in the range, $2\leq\tan{\beta}\leq 60$. The bottom dashed-lines represent the experimental limits from Super-K and top dashed-lines represent Hyper-K limits.}
\label{fig3}
\end{figure}

\begin{figure}[t]\centering
	\subfloat[$p \rightarrow e^+ \pi^0 $]{\includegraphics[width=3.278in]{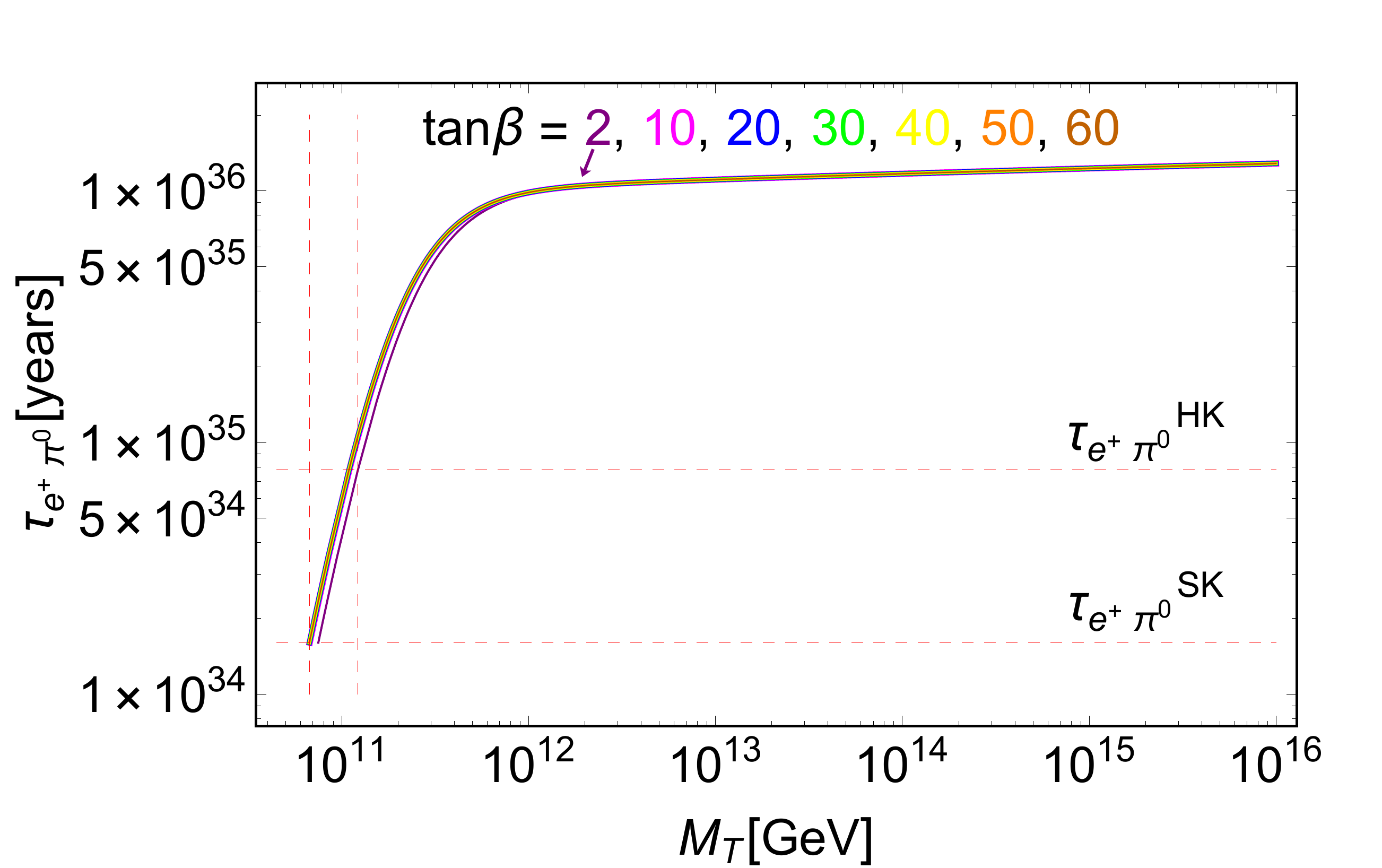}} \hspace{0.1cm}
	\subfloat[$p \rightarrow \mu^+ \pi^0 $]{\includegraphics[width=3.1in]{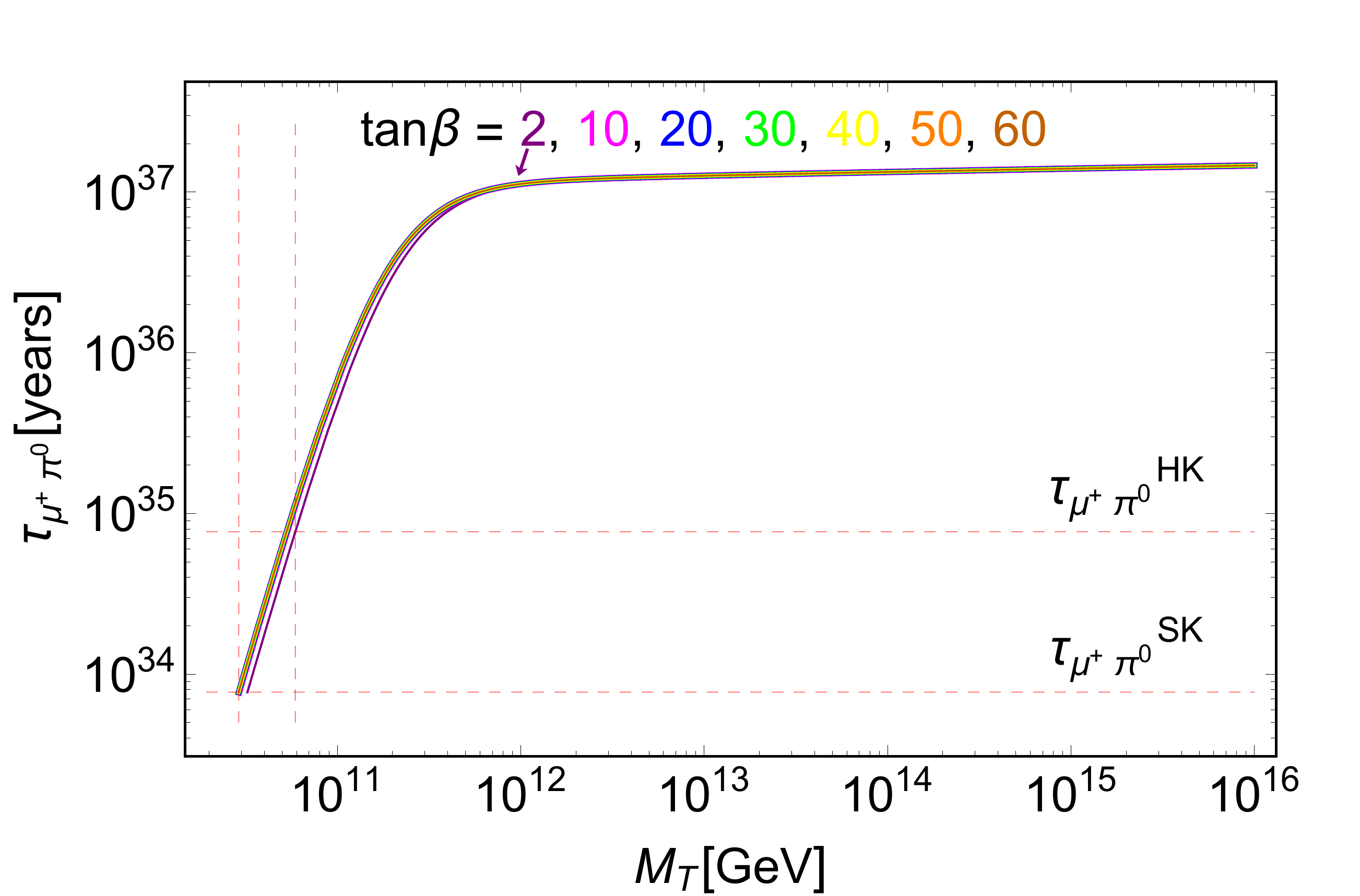}\label{mupi33}}\\
	\subfloat[$p \rightarrow e^+ K^0 $]{\includegraphics[width=3.1in]{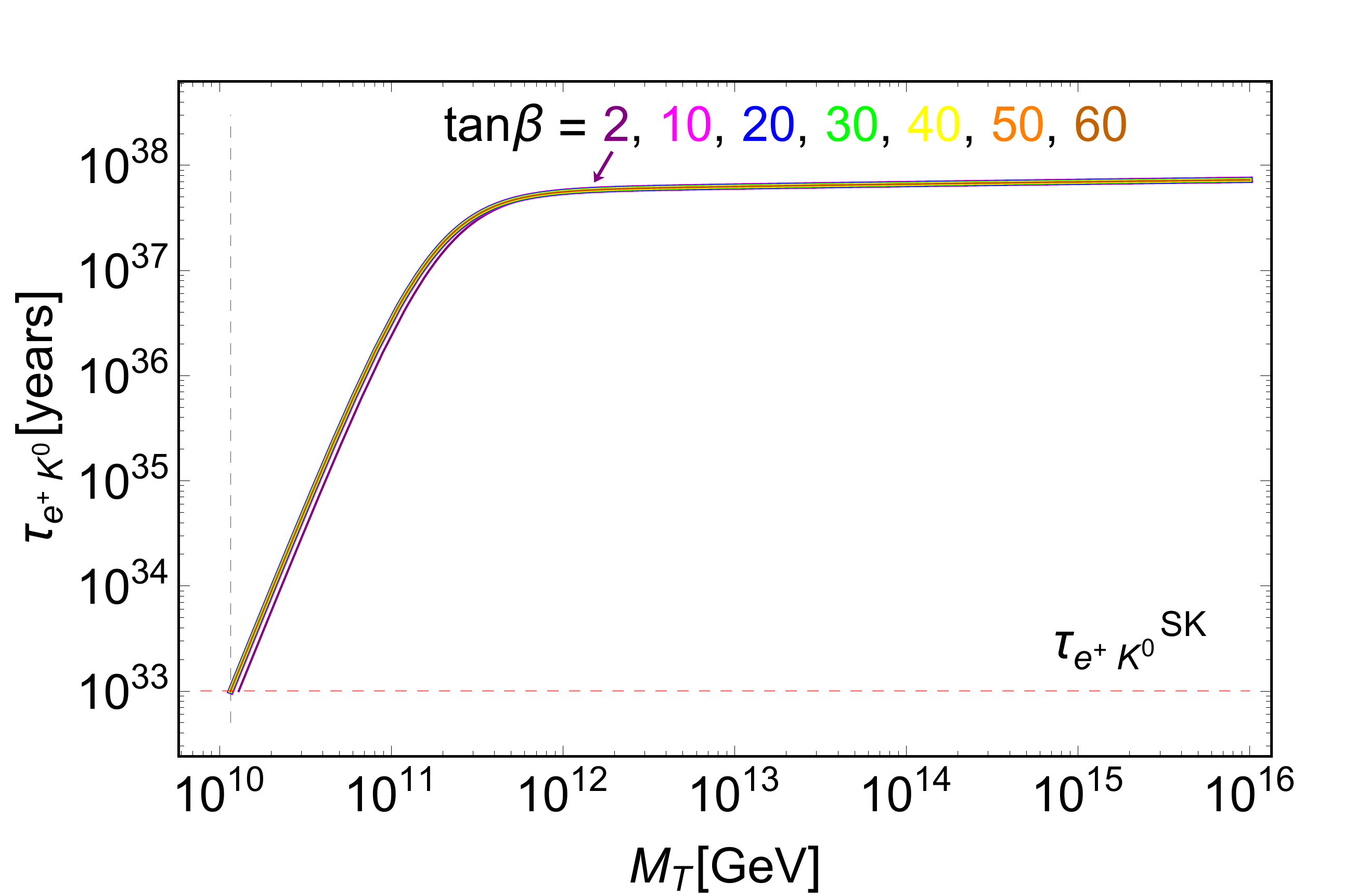}\label{epi34}} \hspace{0.1cm}
	\subfloat[$p \rightarrow \mu^+ K^0 $]{\includegraphics[width=3.1in]{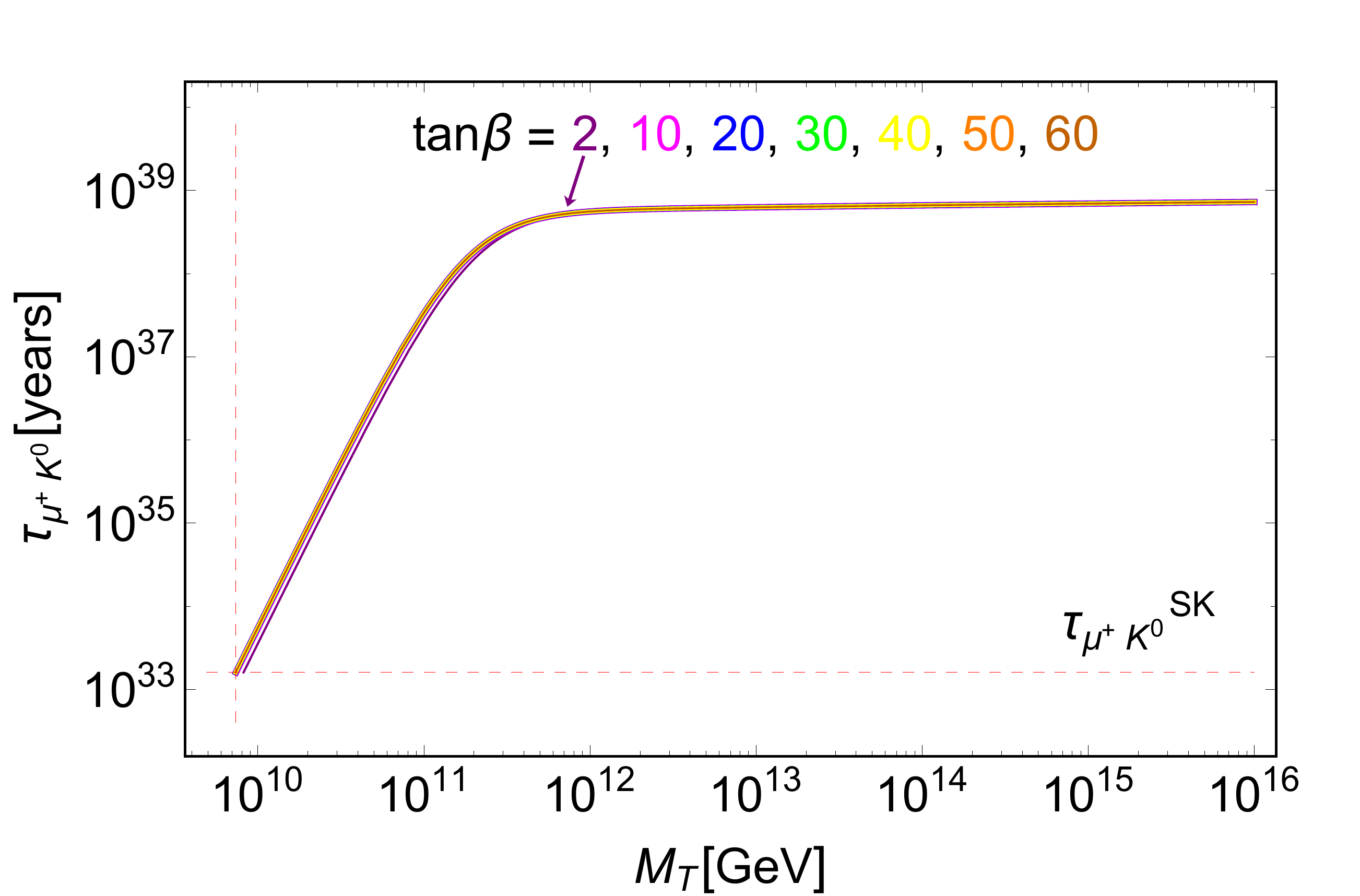}\label{mupi34}}\\
	\caption{Estimates of proton partial lifetime for charged lepton decay channels as a function of 5-plet triplet mass $M_T = M_{\bar{\lambda}} \ll M_{\lambda}$ with $\tan \beta$ in the range $2\leq\tan{\beta}\leq 60$. The lower dashed lines represent the experimental limits from Super-K and the upper dashed lines in (a) and (b) represent future Hyper-K limits.}
\label{fig4}
\end{figure}

The numerical estimates of partial proton lifetime  for charged lepton decay channels are shown in Fig.~(\ref{fig3}) as a function of the triplet mass $M_T = M_{\bar{\lambda}} = M_{{\lambda}}$, with $\tan \beta$ values in the range $2\leq\tan{\beta} \leq 60$. The gauge boson contribution becomes dominant for $\bar{\lambda} = \lambda \gtrsim (\sqrt{m_{(d,s)} m_{(e,\mu)}}/ v_d)$ and for $\bar{\lambda} = \lambda <  (\sqrt{m_{(d,s)} m_{(e,\mu)}}/ v_d)$ the contribution of color triplet with mass $M_{\lambda}$ remains dominant over other contributions. 
We can use the experimental bounds on the various partial proton lifetimes depicted in Table~\ref{matrix-el} to calculate the corresponding lower bounds on the color triplet masses. For example, using the Super-K bound (Table~\ref{matrix-el}) on proton lifetime for the decay channel $p \rightarrow e^+ \ \pi^0$  and Eq.~(\ref{dr1}), we obtain the lower bound,
\begin{eqnarray}
M_{T} \gtrsim   \left( \sqrt{1+\tan^2 \beta} \right) \, 4.56 \times 10^{10} \text{ GeV} \quad \text{or} \quad \overline{\lambda}=\lambda  \gtrsim \left( \sqrt{1+\tan^2 \beta} \right) \, 3.24 \times 10^{-6}. \label{MTbound}
\end{eqnarray}
Thus, the Super-K bound on the decay channel $p \rightarrow e^+ \ \pi^0$ corresponds to $M_T \gtrsim 1.0 \times 10^{11} - 2.7 \times 10^{12} $~GeV or $\overline{\lambda} =\lambda \gtrsim 7.2 \times 10^{-6} - 1.9 \times 10^{-4} $ for $\tan \beta$ in the range $2\leq\tan{\beta}\leq 60$. With an additional $Z_4$ symmetry these relatively tiny values of $\bar{\lambda} = \lambda$ can be boosted by a factor $(m_P/M)^2 \sim 10^4$ as discussed in the subsection below.

As the contributions from both color triplets become comparable for
$\lambda \sim (\tan \beta) (\sqrt{m_{(d,s)} m_{(e,\mu)}}/m_u) \bar{\lambda} \sim ((\tan \beta)/2)(1.4,\,6.4,\,20.6,\,91.8) \bar{\lambda}$, it is instructive to consider the limit $\lambda \gg ((\tan \beta)/2)(1.4,\,6.4,\,20.6,\,91.8) \bar{\lambda}$, where the contribution from the color triplet of mass $M_{\bar{\lambda}}$ becomes dominant over the other contributions for $\overline{\lambda} \lesssim m_u/v_u$. The numerical prediction for this scenario is depicted in Fig.~(\ref{fig4}). As expected from the $M_{\overline{\lambda}}$ contribution in Eqs.~(\ref{dr1}) and (\ref{dr2}), the weak dependence on $\tan \beta$ in the range $2\leq\tan{\beta}\leq 60$ does not exhibit any spread in the proton lifetime predictions shown in Fig.~(\ref{fig4}). In this case the Super-K bound for the decay channel $p \rightarrow e^+ \ \pi^0$ with Eq.~(\ref{dr1}) gives the following lower bound,
\begin{eqnarray}
M_{\overline{\lambda}} \gtrsim 7.4 \times 10^{10} \text{ GeV} \quad \text{with} \quad \overline{\lambda}  \gtrsim  5.2 \times 10^{-5}.
\end{eqnarray}
which is somewhat smaller than the corresponding estimate of $M_{\lambda}$ contribution quoted in Eq.~(\ref{MTbound}). However, a potentially observable range of this bound, with $M_T \lesssim 10^{11}$~GeV, is in contradiction with Super-K bounds on neutral lepton channels described below.

\begin{figure}[t]\centering 
\subfloat[$p \rightarrow \bar{\nu} \pi^+ $]{\includegraphics[width=3.1in]{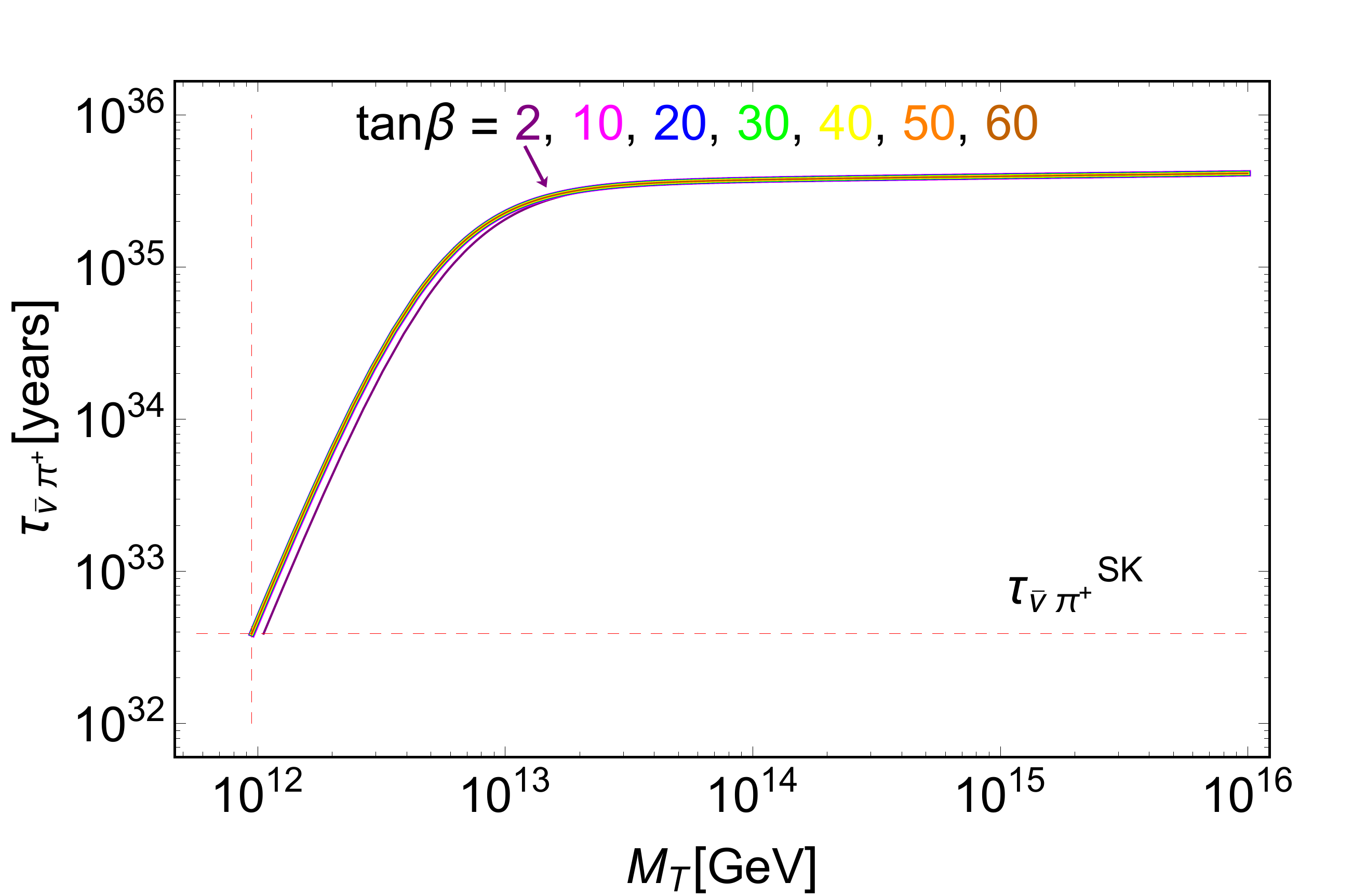}\label{nupi}}\hspace{0.1cm}
\subfloat[$p \rightarrow \bar{\nu} K^+ $]{\includegraphics[width=3.11in]{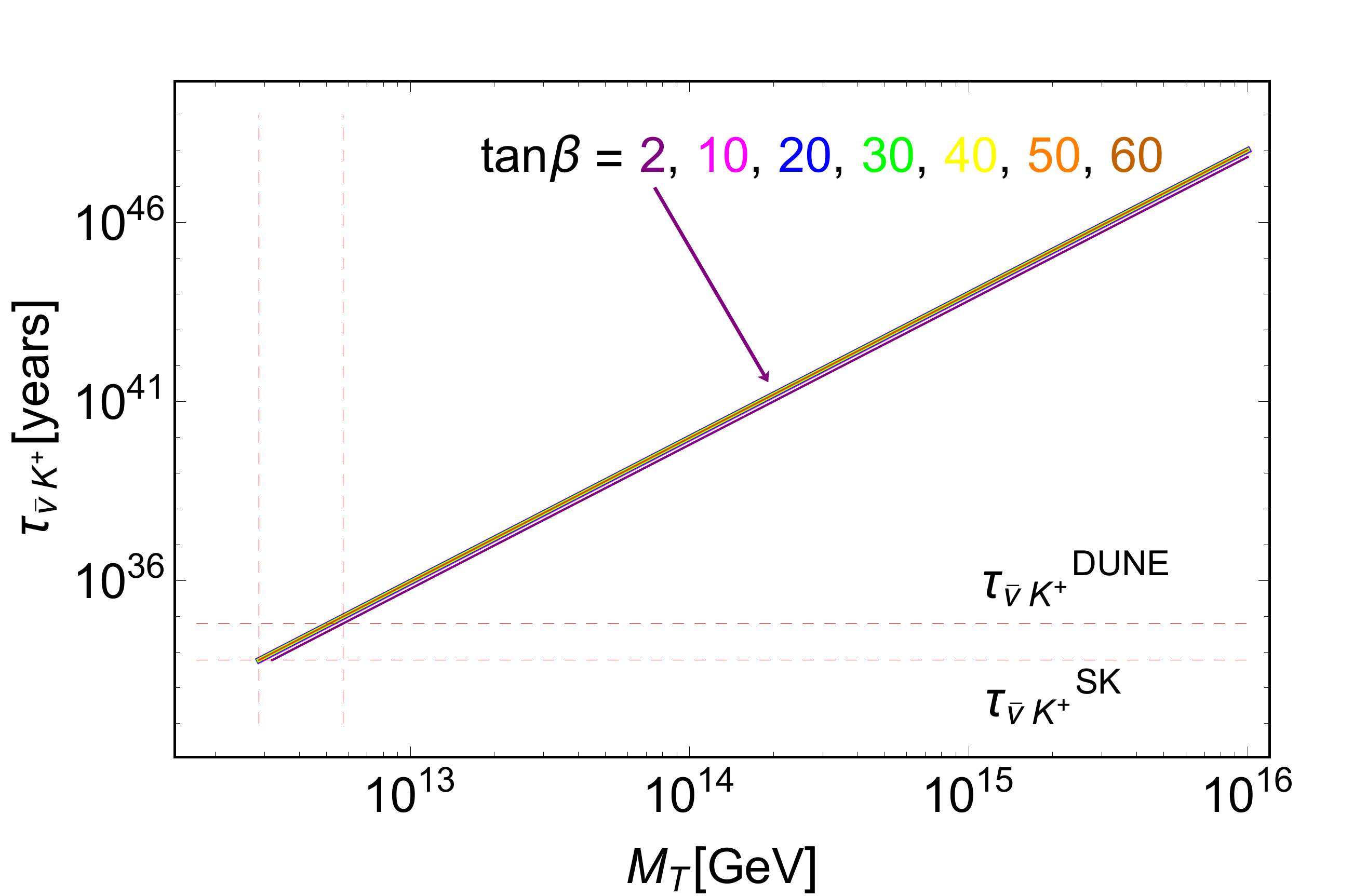}\label{nuk}} 
\caption{Estimates of proton partial lifetime for neutral lepton decay channels as a function of 5-plet triplet mass $M_T = M_{\bar{\lambda}}= M_{{\lambda}}$  with $\tan \beta$ in the range $2\leq\tan{\beta}\leq 60$. The lower dashed-line in (b) represent the experimental limit from Super-K and the upper dashed-line represent future DUNE limit.}
\end{figure}

The proton decay rates for neutral lepton channels, $\pi^+ \bar{\nu}_i$ and $K^+ \bar{\nu}_i$, based on the neutrino model described in $W^I_{HN}$ (Eq.~\ref{WII}), are expressed as  

\begin{eqnarray}
\Gamma_{p \rightarrow  \pi^+ \bar{\nu_i}}& = & k_{\pi}|T_{\pi^+ \bar{\nu}}|^2 A^2_{S_1} \left|\frac{(U^*_N)_{i1}}{M^2} +  V_{ud} \, \frac{m_{u}}{\upsilon_u} \sum_j \frac{(m^{(u)})_j}{\upsilon_u} \frac{ (U^*_N)_{ij} \, (V)_{j1}}{M^2_{\bar{\lambda}}}\right|^2, \label{dr4}
\end{eqnarray}
\begin{eqnarray}
\Gamma_{p \rightarrow  K^+ \bar{\nu_i}}& = & k_{K} A^2_{S_1}\left| e^{i\varphi_1}T^{\prime}_{K^+ \bar{\nu}} (V^*)_{ud} \, \frac{m_{u}}{\upsilon_u} \sum_j \frac{(m^{(u)})_j}{\upsilon_u} \frac{ (U^*_N)_{ij} \, (V)_{j2}}{M^2_{\bar{\lambda}}} \right. \nonumber \\
&+& \left. e^{i\varphi_2}T^{\prime \prime}_{K^+ \bar{\nu}}(V^*)_{us} \, \frac{m_{u}}{\upsilon_u} \sum_j \frac{(m^{(u)})_j}{\upsilon_u} \frac{(U^*_N)_{ij} \, (V)_{j1}}{M^2_{\bar{\lambda}}} \right|^2. \label{dr5}
\end{eqnarray}
Apart from the gauge boson contribution in the $\pi^+ \bar{\nu}_i$ channel \cite{Ellis:2020qad} the contribution of color triplet with mass $M_{\overline{\lambda}}$ has been ignored so far. The numerical results are displayed in Figs.~(\ref{nupi}) and (\ref{nuk}) where we have used the recently updated values of $U_{PMNS}$ parameters from \cite{Esteban:2020cvm} with $U_{N}$ equal to the unit matrix. In the large $M_T$ limit the proton lifetime of the first channel is dominated by the gauge boson contribution  whereas for the second decay channel lifetime increases without bound  due to the absence of the gauge boson contribution. 

For neutral lepton channels the Super-K bound for the decay channel $p \rightarrow \overline{\nu} \ K^+ $ with Eq.~(\ref{dr5}) gives the following lower bound,
\begin{eqnarray}
M_{\overline{\lambda}} \gtrsim 3.2 \times 10^{12} \text{ GeV} \quad \text{with} \quad \overline{\lambda}  \gtrsim  2.2 \times 10^{-4}.
\end{eqnarray}
This is the largest bound among the neutral and charged lepton channels with a naturally accessible value with $Z_4$ symmetry. This bound also allows the charged lepton channels, shown in Fig.~(\ref{fig3}) with $\overline{\lambda} \gtrsim 2.2 \times 10^{-4} > \lambda  \gtrsim \left( \sqrt{1+\tan^2 \beta} \right) \, 3.24 \times 10^{-6}$, to lie within the observable range of Hyper-K whereas the prediction of $\pi^+ \bar{\nu}$ channel lies far beyond the reach of future detectors.  

It is important to note that DUNE is more sensitive to the $p \rightarrow K^+ \bar{\nu}$ channel than Hyper-K \cite{Acciarri:2015uup}. This is mainly due to fact that the DUNE collaboration will exploit liquid Argon time-projection chamber technology which can identify $K^+$ track with high efficiency as compared to a water Cherenkov detector like Super-K or Hyper-K. However, JUNO, which is a liquid scintillator detector \cite{An:2015jdp} is expected to provide us the world's best limit for $p \rightarrow K^+ \bar{\nu}$ mode by the mid-2020s, before DUNE/Hyper-K catch up with the JUNO’s limit.

\begin{figure}[ht!]\centering
	\subfloat[]{\includegraphics[width=3.15in]{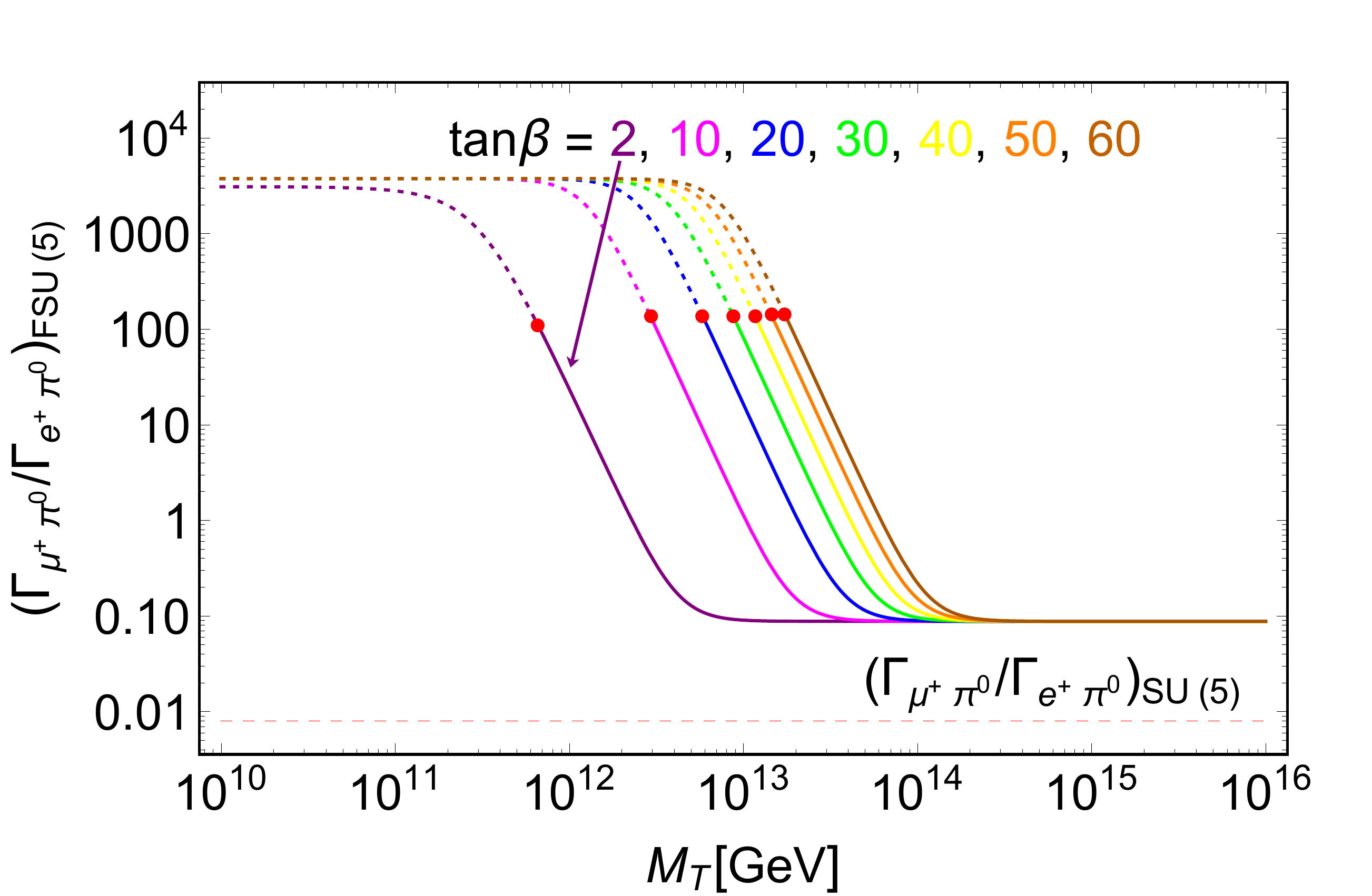}} \hspace{0.1cm}
	\subfloat[]{\includegraphics[width=3.15in]{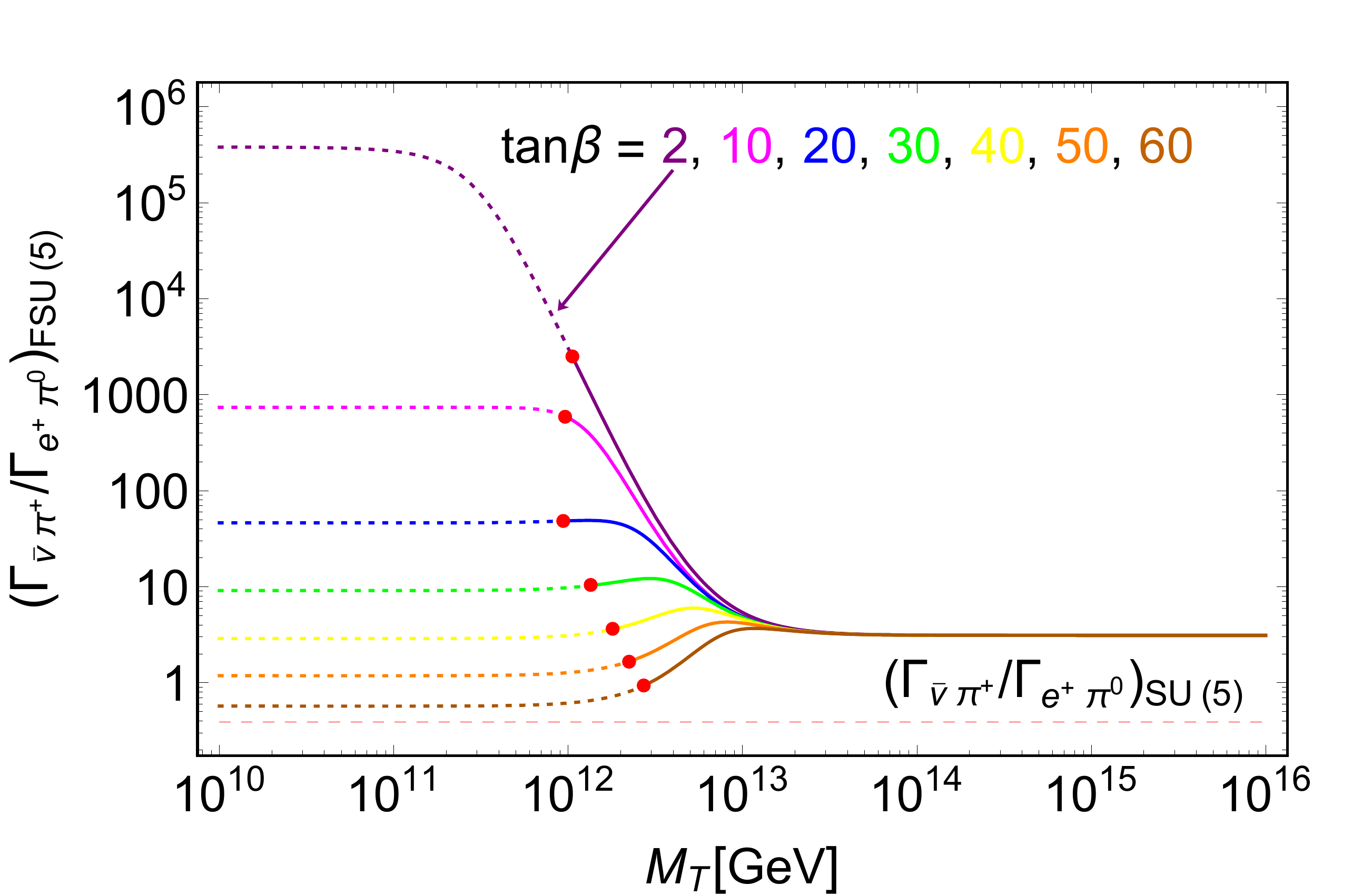}}\\
	\subfloat[]{\includegraphics[width=3.16in]{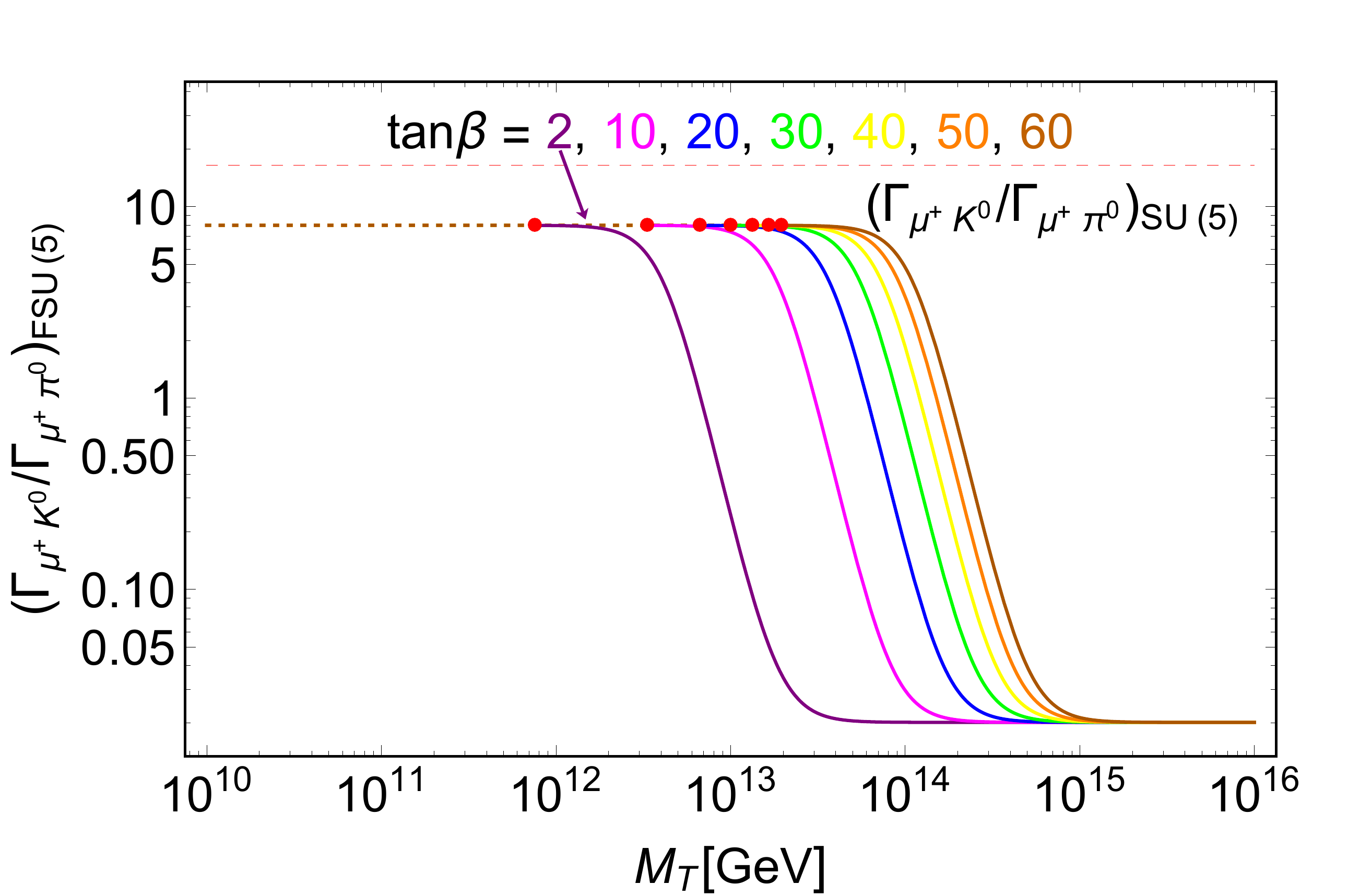}} \hspace{0.1cm}
	\subfloat[]{\includegraphics[width=3.15in]{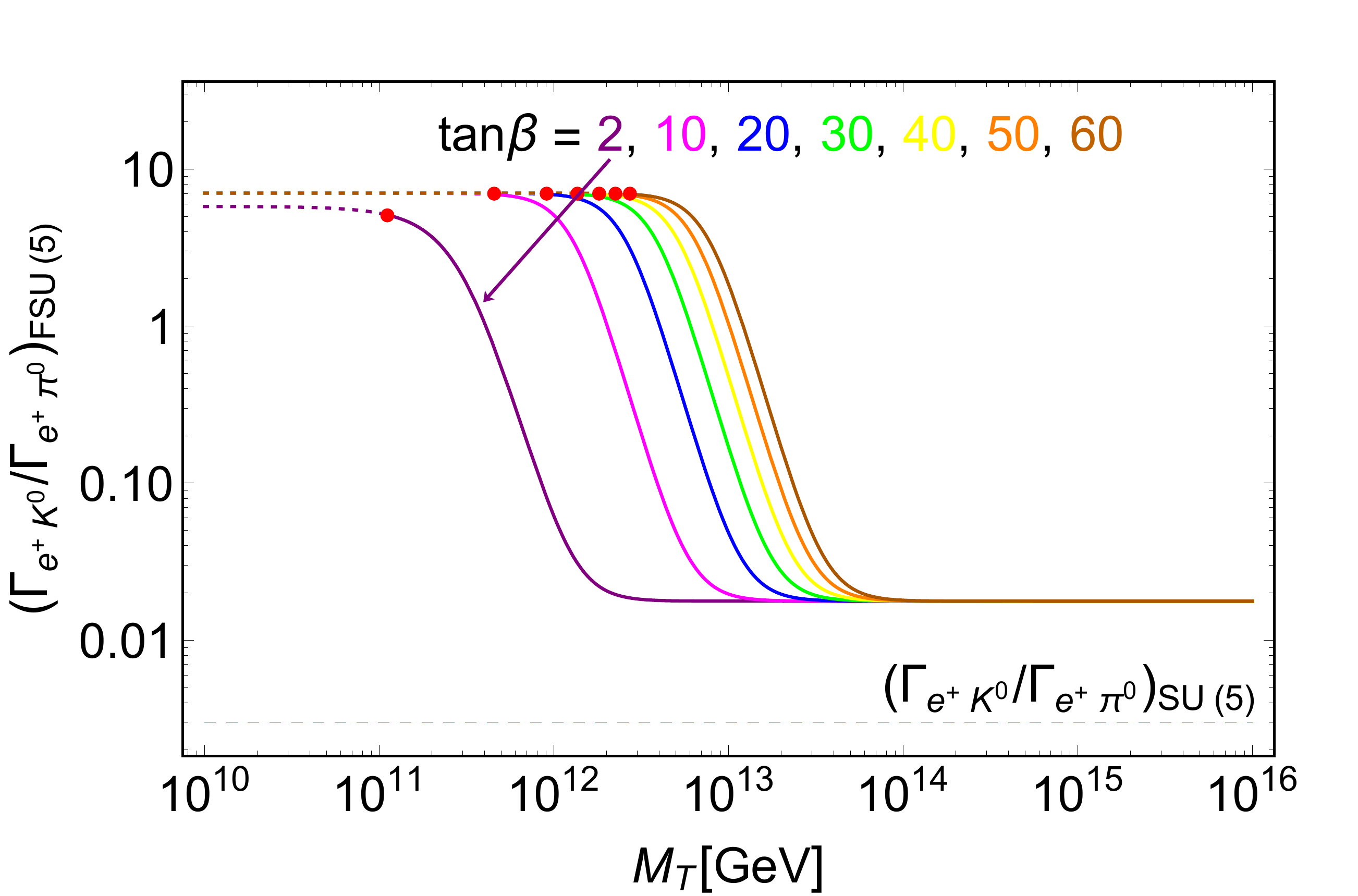}}\\
	\subfloat[]{\includegraphics[width=3.15in]{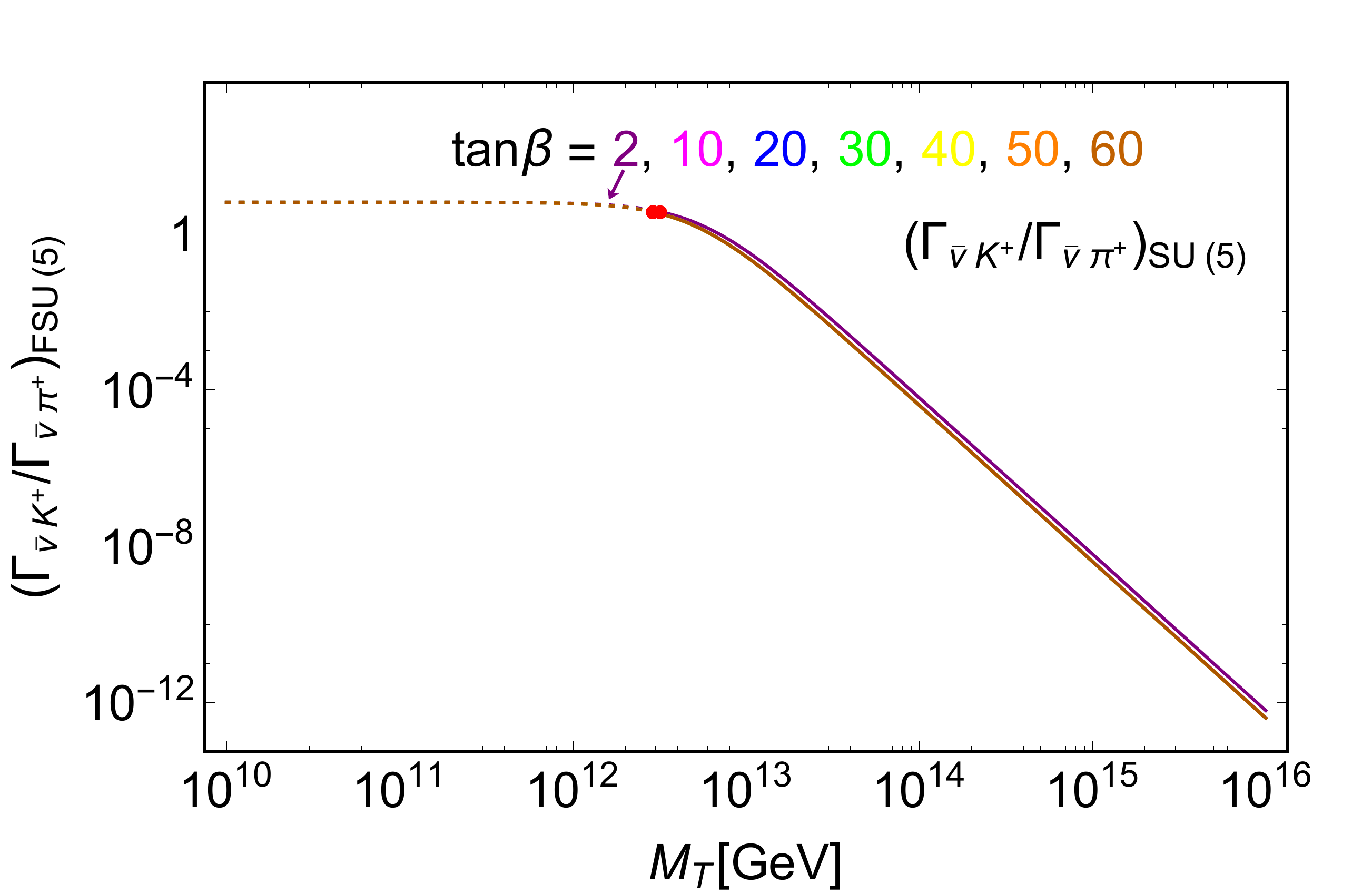}}\hspace{0.1cm}
	\subfloat[]{\includegraphics[width=3.15in]{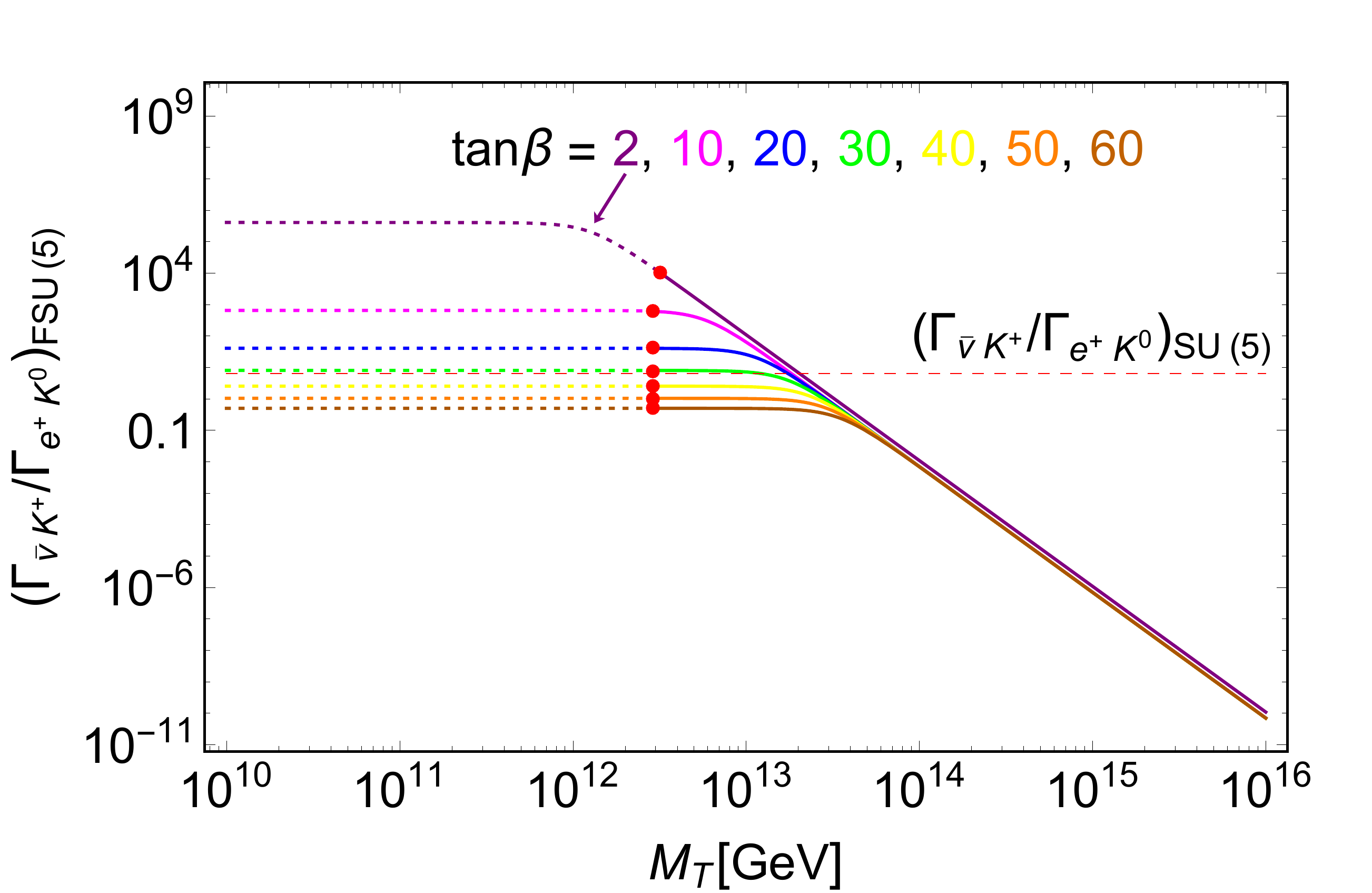}}
	\caption{Estimates of various branching fractions as a function of triplet mass $M_T = M_{\bar{\lambda}}= M_{{\lambda}}$,  with $\tan \beta$ in the range $2\leq\tan{\beta}\leq 60$. For comparison the corresponding predicted values of branching fraction for $SU(5)$ GUT are included from  \cite{Ellis:2020qad}. The solid line curves of $FSU(5)$ predictions are consistent with the Super-K bounds shown with  red dots.}
	\label{fig6}
\end{figure}

In order to make a comparison of proton partial lifetime predictions among various GUT models the estimates of branching fractions play a pivotal role. For this purpose a variation of various branching fractions with respect to color triplet mass $M_T = M_{\bar{\lambda}}= M_{{\lambda}}$ for 
$\tan \beta$ in the range $2\leq\tan{\beta}\leq 60$ is shown in Fig.~(\ref{fig6}). We particularly include the corresponding predictions from the unflipped $SU(5)$ model recently presented in \cite{Ellis:2020qad} by ignoring the dimension five contribution of color triplets with large sfermion masses of order $100$~TeV \cite{Hisano:2013exa,Nagata:2013sba,Nagata:2013ive,Evans:2015bxa,Ellis:2015rya,Ellis:2016tjc,Evans:2019oyw,Ellis:2019fwf}. For a comparison with  4-2-2 model see \cite{Lazarides:2020bgy} and for $SO(10)$ models see Refs.~\cite{Babu:1997js,Babu:1998wi,Haba:2020bls}. As is obvious from Fig.~(\ref{fig6}) the present $FSU(5)$ model makes a very distinctive predictions of various branching fractions within the observable range of Hyper-K. Especially the branching fraction of $\overline{\nu} \ K^+ $ channel plays a key role in making distinctive  comparison of the current model with the other models of flipped $SU(5)$ \cite{Ellis:2020qad,Hamaguchi:2020tet} where this channel is highly suppressed.

\subsection{$Z_4$ Symmetry and Color Triplet Masses}
An additional $Z_4$ symmetry can be employed to make the color triplets naturally light for observable proton decay. This is achieved with the following $Z_4$-charge assignments:
\begin{equation}
q_{Z_4}(10_H,\,\overline{10}_H,\,\mathcal{S}_a) \rightarrow (1,\,1,\,1),
\end{equation} 
with all other fields carrying zero $Z_4$-charge. This modifies the superpotential in Eq.~(\ref{sp1}) as follows:

\begin{align} \label{sp11}
W_{Z_4} &=\kappa S \left( \frac{\left(10^1_{H}\overline{10}^{-1}_{H}\right)^2}{m_P^2}  - M^2 \right) 
\nonumber \\
& + \frac{\lambda}{16} \left(\frac{10^1_{H}\overline{10}^{-1}_{H}}{m_P^2} \right) 10^1_{H}10^1_{H}5^{-2}_{h} + \frac{\overline{\lambda}}{16} \left(\frac{10^1_{H}\overline{10}^{-1}_{H}}{m_P^2} \right) \overline{10}^{-1}_{H} \overline{10}^{-1}_{H} \overline{5}^2_{h} \nonumber 
\\
& + \frac{1}{8} \,  y^{(d)}_{ij}10^1_{i}10^1_{j}5^{-2}_{h}+y^{(u,\nu)}_{ij}10^1_{i}\overline{5}^{-3}_{j}\overline{5}^{2}_{h}+y^{(e)}_{ij}1^5_{i}\overline{5}^{-3}_{j}5^{-2}_{h}+\frac{\gamma_{ai}}{4} \left(\frac{10^1_{H}\overline{10}^{-1}_{H}}{m_P^2} \right) \mathcal{S}_a 10^1_i \overline{10}^1_H.
\end{align}
This superpotential can be employed to realize smooth hybrid inflation \cite{Rehman:2012gd}. Also see \cite{Rehman:2018gnr} for a relevant model of inflation. The $Z_4$ symmetry is spontaneously broken during smooth hybrid inflation and the domain wall problem is therefore avoided.

It is important to note that both color triplets are now  naturally light relative to $M_G$ with $M_{\lambda}=\lambda M (M/m_P)^2$ and $M_{\overline{\lambda}}=\overline{\lambda} M (M/m_P)^2$, and the couplings, $\gamma_{ai}$, relevant for the realization of light neutrino masses via a double seesaw mechanism have also been enhanced by the factor, $(m_P/M)^2$. The explicit mass term, $\mu_{ab} \mathcal{S}_a \mathcal{S}_b$, for the gauge singlet fields $\mathcal{S}_a$, generated effectively from the K\"ahler potential, $K\supset y_{ab} \frac{\Sigma^{\dagger}}{m_P} \mathcal{S}_a \mathcal{S}_b + h.c$, still remains intact. Note that we do not consider this symmetry in the second model of neutrino masses based on the standard seesaw mechanism arising from the explicit $R$-symmetry breaking terms at nonrenormalizable level.

\section{$\bm{U(1)_{R}}$ Violation and proton decay}\label{U1R1SB}
\begin{figure}[t]\centering
\subfloat[]{\includegraphics[width=2.0in]{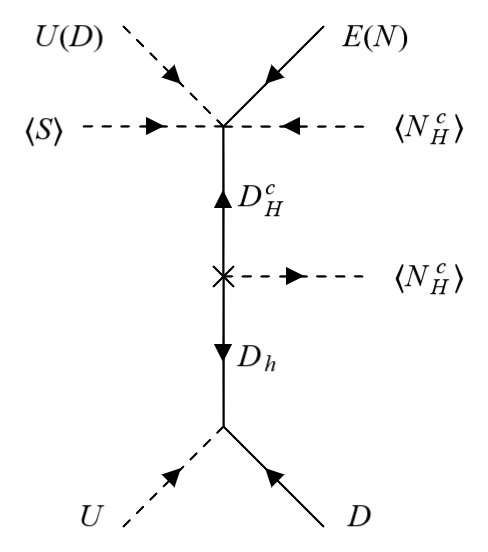}} \hspace{2.0cm}
\subfloat[]{\includegraphics[width=2.0in]{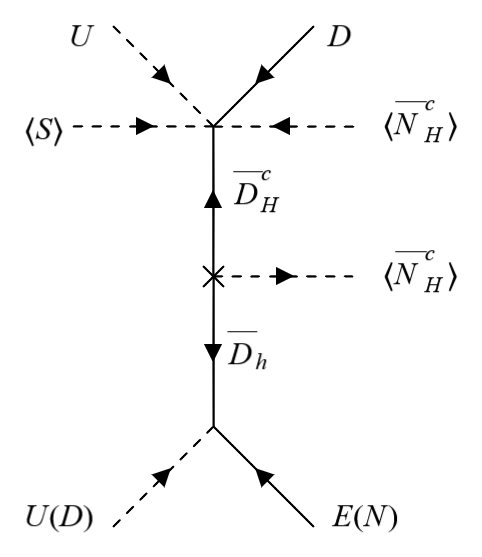}} \\
\subfloat[]{\includegraphics[width=2.0in]{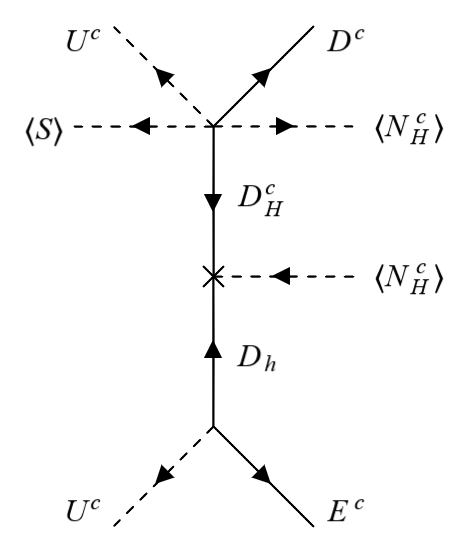}} \hspace{2.0cm}
\subfloat[]{\includegraphics[width=2.0in]{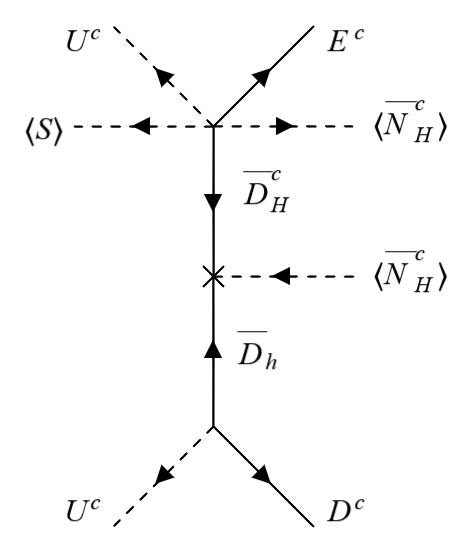}}
\caption{Dimension five proton decay diagrams corresponding to effective operators $10\,10\,10\,\overline{5}$ and $10\,\overline{5}\, \overline{5}\,1$ mediated via color triplets $(D_H^c,\,D_h) \subset (10_H,\,5_h)$ and $(\overline{D_H^c},\, \overline{D_h}) \subset (\overline{10}_H,\,\overline{5}_h)$ are shown in panels (a),(c) and (b),(d) respectively. The dashed (solid) lines represent bosons (fermions).\label{510}}
\end{figure}

In this section we first discuss the effect of $U(1)_{R}$ symmetry breaking coming from the soft susy breaking terms. For example, consider the following nonrenormalizable terms allowed by the $R$-symmetry,
\begin{eqnarray}
W &\supset & 
 \frac{\eta_1}{8}\frac{S}{m_P} \left(\frac{ 10^1_H 10^1_H 10^1 \overline{5}^{-3}}{m_P}\right) + \frac{\eta_2}{4} \frac{S}{m_P} \left(\frac{ {(10^1 \overline{10}^{-1}_H}) \cdot ( 10^1 \overline{10}^{-1}_H)}{m_P}\right)  \nonumber \\
&+&  \frac{\eta_3}{8} \frac{S}{m_P} \left(\frac{ {\overline{10}^{-1}_H}\overline{10}^{-1}_H\ \overline{5}^{-3} 1^5}{m_P}\right).
\end{eqnarray}
Once $S$ attains a nonzero vev, $\langle S \rangle \sim \kappa\,m_{3/2}$, from the soft susy breaking terms \cite{Dvali:1997uq}, these interactions can lead to dimension five proton decay diagrams, as shown in Fig~.(\ref{510}), from mixing of the color-triplets from the Higgs 5-plets and 10-plets. However, the presence of a small factor $(\kappa m_{3/2}/m_P) \lesssim 10^{-12}$ renders such decays from operators of type $LLLL$ and $RRRR$ relatively suppressed in comparison to proton decay operators of type $LLRR$ discussed earlier. The suppression of these dimension five operators is a distinctive feature of the present $R$-symmetric model described in Eq.~\ref{sp1}. This feature is absent in the models considered in  \cite{Ellis:2020qad} and \cite{Hamaguchi:2020tet} where the above mentioned potentially dangerous dimension five proton decay operators can lead to rapid proton decay with natural values of the couplings involved.

Finally, we briefly comment on the second model of neutrino masses described in the superpotential $W_{HN}^{II}$ (Eq.~(\ref{WII})), where an explicit $R$-symmetry breaking allowing only operators of zero $R$-charge is assumed at the nonrenormalizable level. The effective Yukawa terms in the superpotential $W_{HN}^{II}$ (Eq.~(\ref{WII})), relevant for proton decay mediated by the color triplets $(D_H^c,\,\overline{D_H^c})$ from $(10_H,\overline{10}_H)$, can be expressed in terms of mass eigenstates as
\begin{eqnarray}
W^{II}_{HN} &\supset &  \left(\frac{M}{m_P} \right) ( L \left( U_L \, \gamma_3 \right) Q\, D_H^c +  U^c \left( \gamma_3 V P^* \right) D^c  D_H^c   
\nonumber \\
&-&  \frac{1}{2} Q \left( \gamma_2 \right) Q \, \overline{D_H^c}  +  U^c \left(\gamma_4 \, U_{E^c} \right) E^c \overline{D_H^c} ). \label{WII2}
\end{eqnarray}
The operators in Eq.~(\ref{WII2}) lead to both dimension five and dimension six proton decay mediated by the color triplets $(D_H^c,\,\overline{D_H^c})$ of $(10_H,\overline{10}_H)$ in a  way similar to the recent discussion for 4-2-2 model in \cite{Lazarides:2020bgy}. However, in the present flipped $SU(5)$ model, the diagrams of Fig.~(\ref{510}) make the dominant contribution due to the absence of a suppression factor $\langle S\rangle/m_P$. To avoid rapid dimension five proton decay  from the mixing of the color triplets in the Higgs 5 and 10-plets, we obtain the following  order of magnitude estimate for the upper bound on $\gamma_{2,3,4}$ couplings,
\begin{eqnarray}
\gamma_{2,3,4} & \lesssim &  10^{-5}.
\end{eqnarray}
Here we use  the experimental bound, $\tau_{\bar{\nu} K^+} \gtrsim  5 \times 10^{33}$~years, with $\tan{\beta}\sim10$, $m_{3/2}\sim 10$~TeV and $\lambda = \overline{\lambda} \sim 1$. A similar order of suppression is also expected in the $FSU(5)$ models considered in \cite{Ellis:2020qad,Hamaguchi:2020tet}.

\section{Gauge Coupling Unification}
As emphasized in section~\ref{U1R1} with the color triplets of intermediate mass the proton lifetime is predicted to lie within the observable range. This brings forth the issue of gauge coupling unification which is otherwise achieved naturally  with the MSSM matter content. In $FSU(5)$ the two MSSM gauge couplings, $g_2$ and $g_3$, unify with the $SU(5)$ gauge coupling $g_5$ at the scale, $M_{23} = g_5\,M$. The third MSSM gauge coupling $g_1$ and the $U(1)_X$ gauge coupling $g_X$ are  related to $g_5$ at $M_{23}$ as
\begin{equation}
\frac{25}{g_1^2(M_{23})} = \frac{1}{g_X^2(M_{23})} + \frac{24}{g_5^2(M_{23})}.
\end{equation}
With a single color triplet pair of mass $\sim 10^{10}-10^{13}$~GeV, the $g_2$ and $g_3$ couplings are unified around $10^{17}-10^{18}$~GeV with $g_1 > g_5$. With two pairs of color triplets the $M_{23}$ unification scale goes beyond the Planck scale. For a possible GUT model beyond $FSU(5)$ based on a simple gauge group we ultimately require $g_5=g_X$ at some scale $M_G$ lying between $M_{23}$ and the string scale $\sim 5\times 10^{17}$~GeV. Therefore, we  need to take care of reconciling  potentially observable proton decay and gauge coupling unification.

To remedy the above mentioned problem we first make a simple choice for the unification scale, $M_{23}=M_G$, with $g_X(M_G)=g_5(M_G)$. This implies that $g_1(M_G)=g_2(M_G)=g_3(M_G)$, which is an attractive feature of MSSM. Next we consider vectorlike 5-plets, $5^{-2}_v+\overline{5}^2_v$, and 10-plets, $10^{1}_v + \overline{10}^{-1}_v$, with odd matter parity and $R(5_v)=R(\overline{5}_v)=1/2$, $R(10_v)=R(\overline{10_v})=1/2$. The relevant superpotential terms for these additional multiplets are
\begin{equation}
W \supset M_5 5^{-2}_v\overline{5}^2_v +
M_{10} 10^{1}_v \overline{10}^{-1}_v + \frac{\lambda_1}{4} 10^1_H 10^{1}_v 5^{-2}_v + \frac{\lambda_2}{4} \, \overline{10}_H^{-1} \overline{10}^{-1}_v \overline{5}^2_v.
\end{equation}
With $Z_4$ symmetry a common mass $M_5$ of the additional vectorlike 5-plets can be achieved around $\sim 10^{12}$~GeV with the help of the suppression factor $(M/m_P)^2 \sim 10^{-4}$ obtained by the $Z_4$ charge assignments, $q_{Z_4}(5_v,\,\overline{5}_v,\,10_v,\,\overline{10}_v)=(1,1,2,2)$. With $M_T \sim 10^{12}$~GeV, we effectively obtain vectorlike 5-plets of intermediate mass with the same number of light doublets and color triplets and this automatically guarantees gauge coupling unification. 

\begin{figure}[ht!]\centering
	\subfloat[]{\includegraphics[width=3.1in]{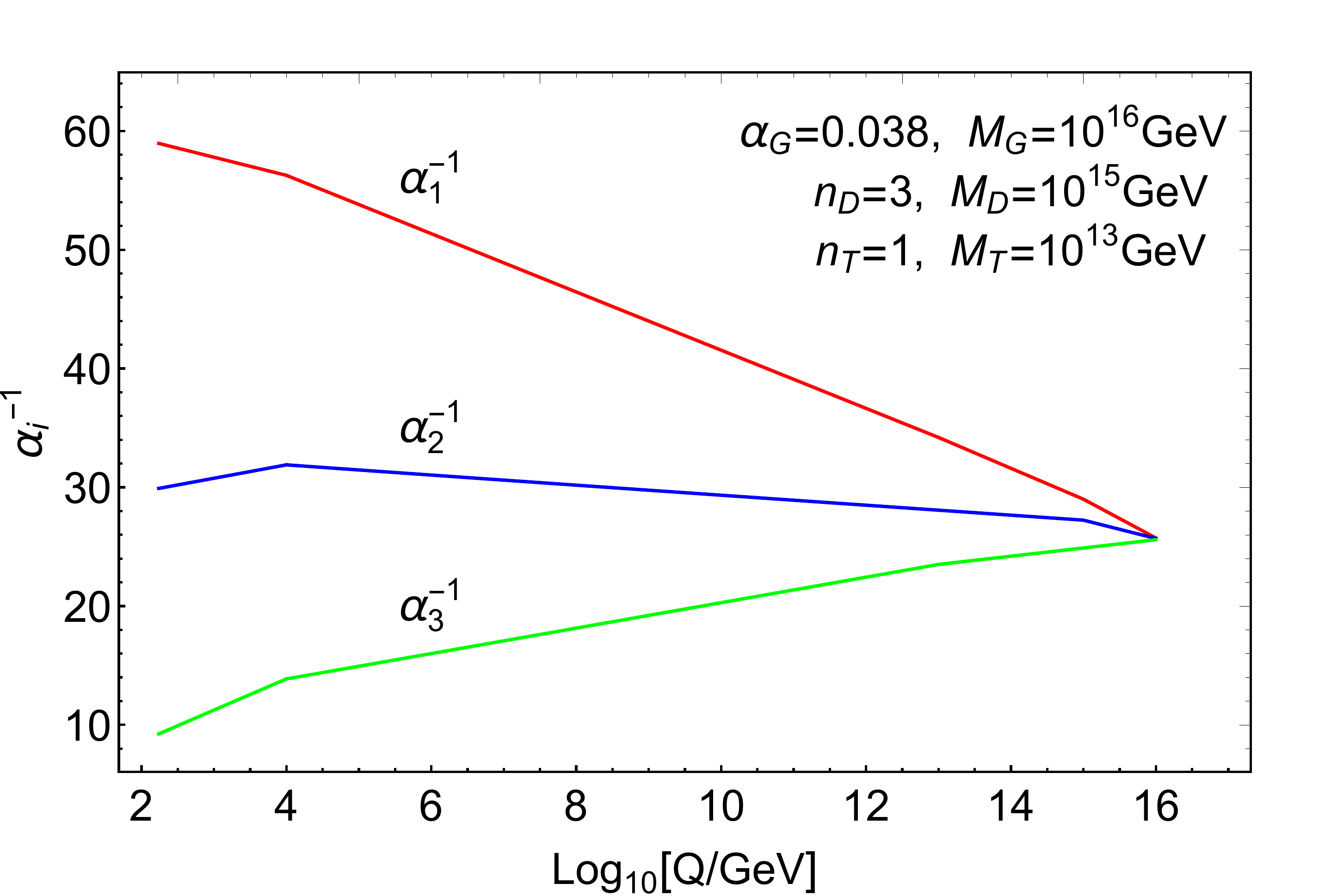}\label{gcu1}}\hspace{0.5cm}
	\subfloat[]{\includegraphics[width=3.1in]{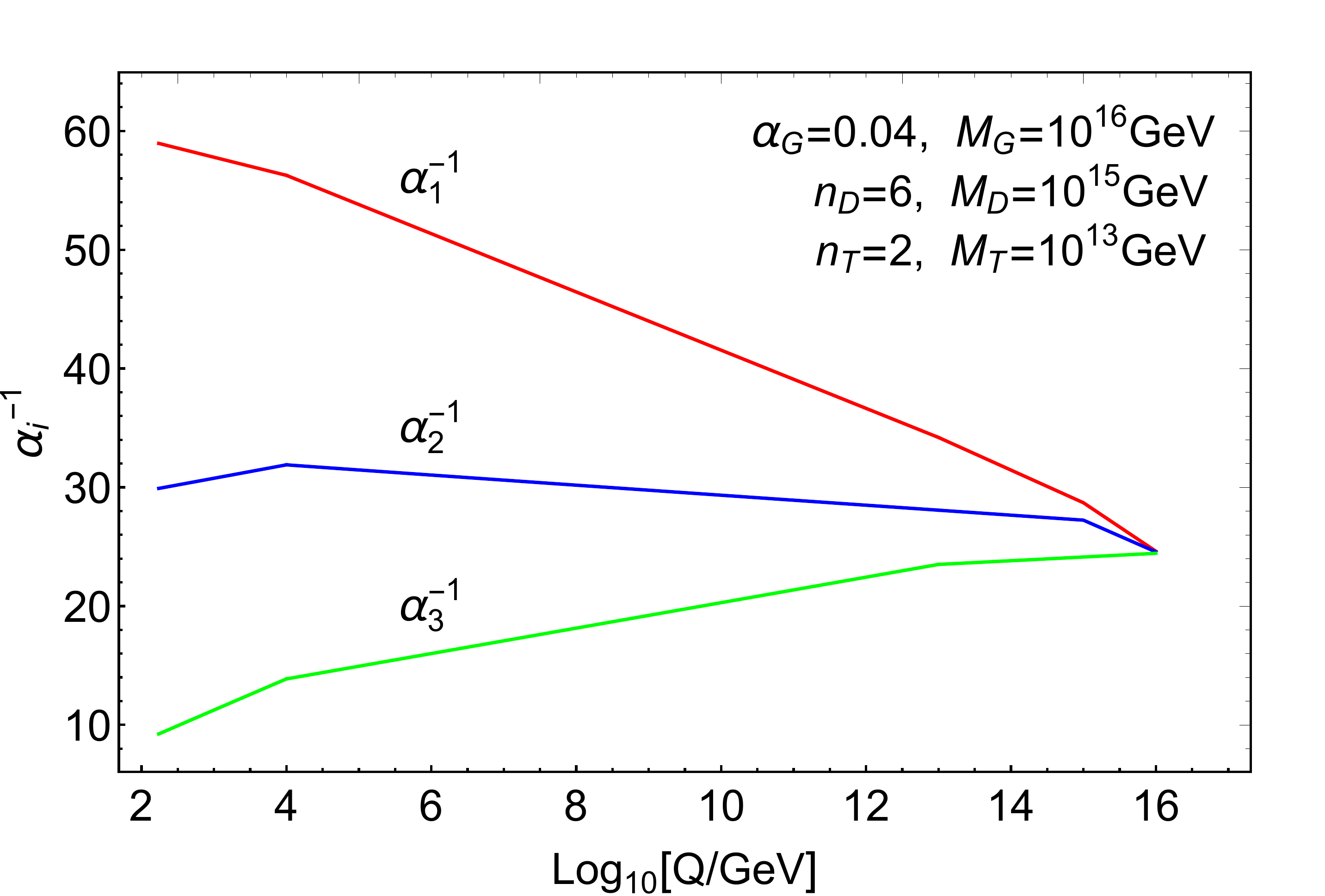}\label{gcu2}}\\
	\subfloat[]{\includegraphics[width=3.1in]{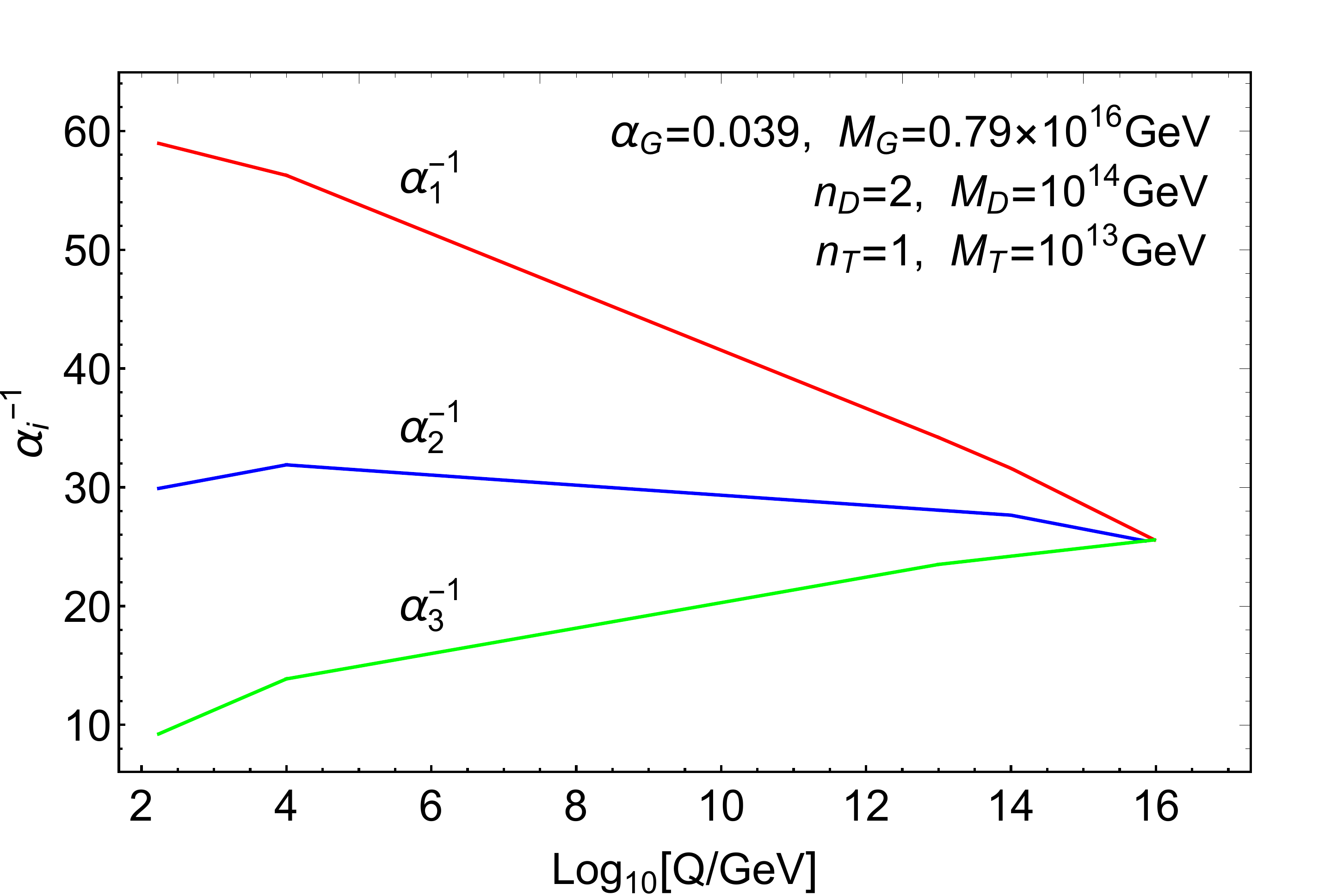}\label{gcu4}} \hspace{0.5cm}
	\subfloat[]{\includegraphics[width=3.1in]{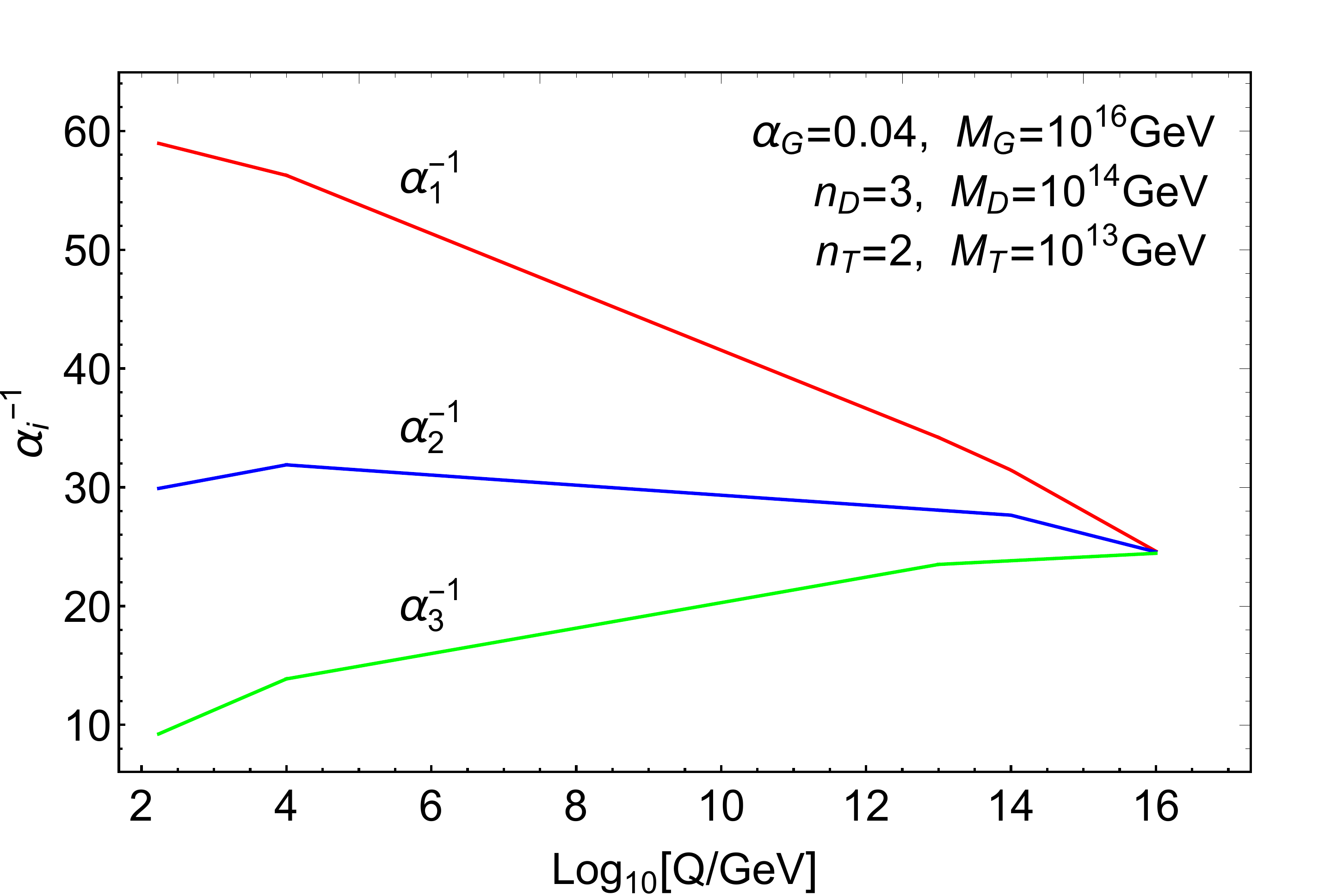}\label{gcu3}}\\
	\subfloat[]{\includegraphics[width=3.1in]{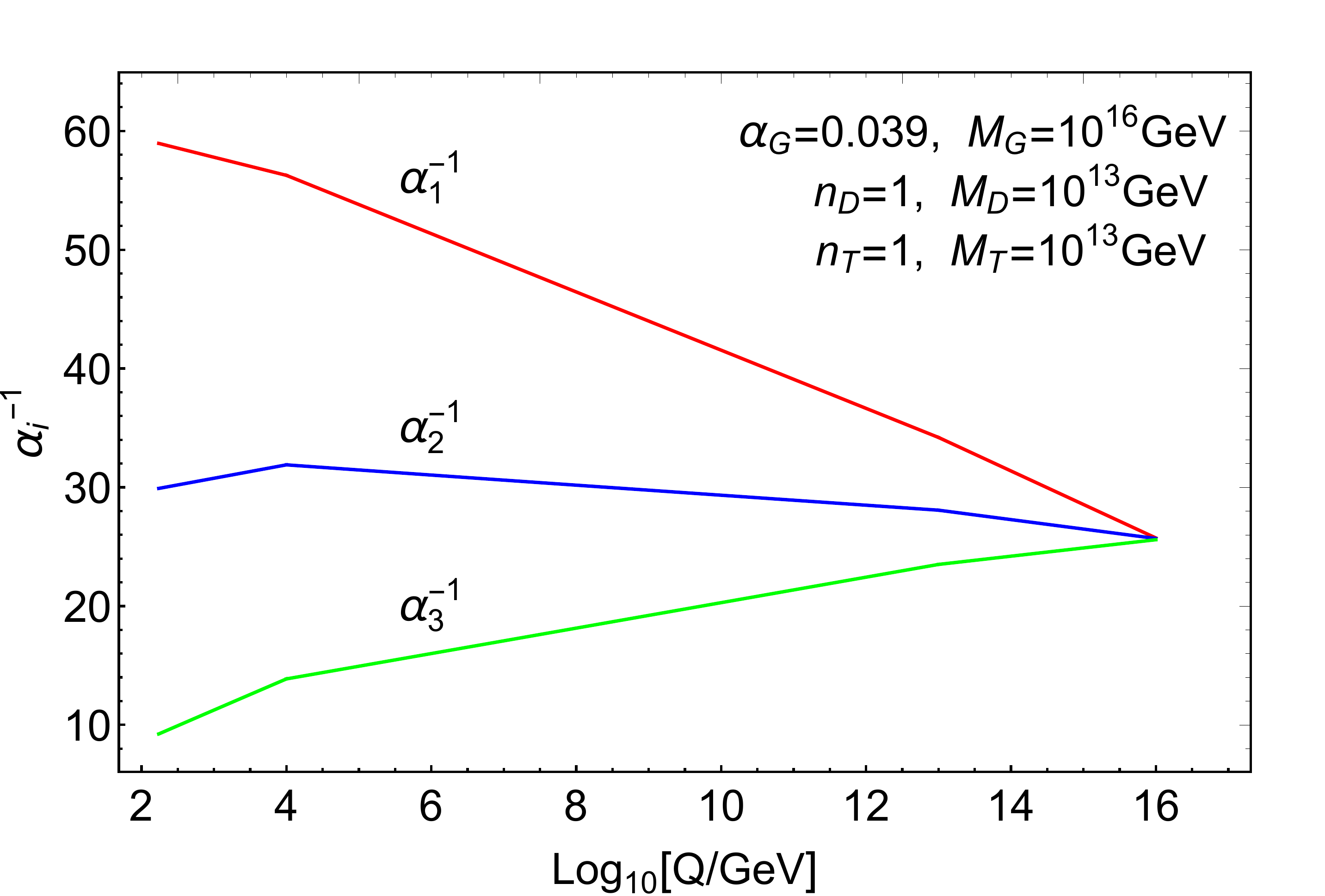}\label{gcu5}} \hspace{0.5cm}
	\subfloat[]{\includegraphics[width=3.1in]{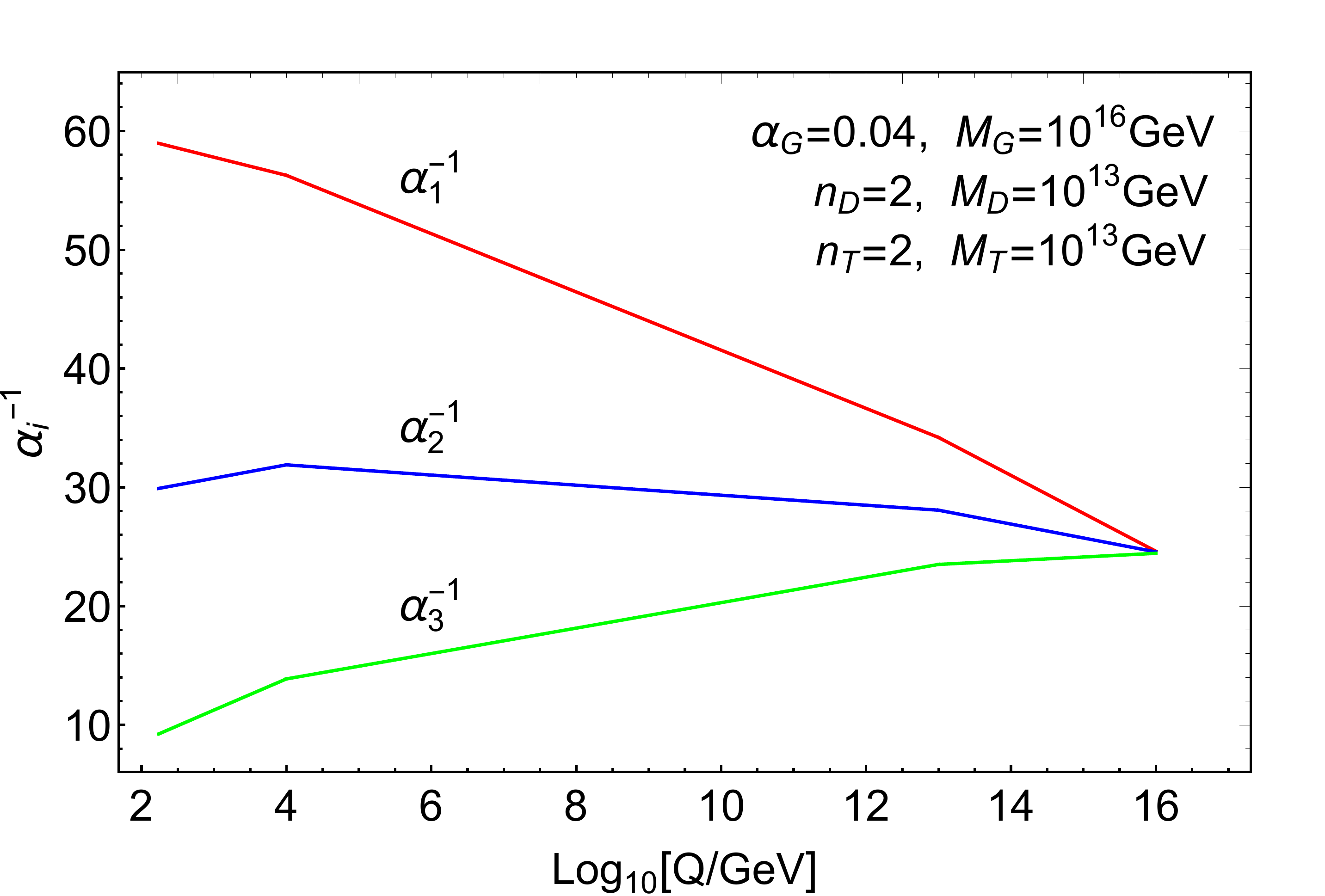}\label{gcu6}}
	\caption{The evolution of inverse gauge couplings $\alpha^{-1}_{i}=4\pi/g^2_i$ versus the energy scale $Q$ with $n_T$ color triplet pairs of intermediate scale mass $M_T=10^{13}$~GeV and $n_D$ additional light doublets with common mass $M_D=(10^{13},\,10^{14},\,10^{15})$~GeV and $M_{\text{SUSY}}=10$~TeV. The values of the unified gauge coupling $\alpha_{G}=g^2/4\pi_G$ at the unification scale $M_G$ are also displayed in all panels.}
	\label{gcu}
\end{figure}

For a scenario without $Z_4$ symmetry we can assume  a common mass $M_5$ of the vectorlike electroweak doublets to lie below the GUT scale, for example within the range, $10^{13}-10^{15}$~GeV. With natural values of the $\lambda_{1,2}$ couplings the color triplets in the additional vectorlike 5-plets, however, achieve GUT scale masses via the missing partner mechanism.
This can lead to gauge coupling unification as shown in Fig.~(\ref{gcu}) for one or two light color triplets with a typical mass value of order $10^{13}$~GeV. With one light color triplet we need two (three) vectorlike doublets at a mass scale $\sim 10^{14}$~GeV ($10^{15}$~GeV), whereas for two light color triplets we require three (six) vectorlike doublets at mass scale $10^{14}$~GeV ($10^{15}$~GeV). In proton lifetime estimates, for simplicity, we assume same mass values for both color triplets and electroweak doublets with $n_T=n_D$. The gauge coupling unification plots for this case are shown  in Fig.~(\ref{gcu5})  for $n_T=n_D=1$ and  Fig.~(\ref{gcu6}) for $n_T=n_D=2$.

\section{\label{con}Conclusion}
Proton decay with lifetimes accessible at Hyper-K and other future experiments are explored in a flipped $SU(5)$ model of supersymmetric hybrid inflation supplemented by a global $U(1)_R$ symmetry. Two distinct models of neutrino masses with normal hierarchy are briefly discussed. Here we discuss the $R$-symmetric model  where the color triplets of intermediate mass from the Higgs 5-plets mediate proton decay with lifetime in the observable range of future experiments. Rapid proton decay mediated through mixing between the color triplets in 5 and 10-plet Higgses severely constrains the relevant couplings in the second neutrino mass model. This decay is adequately suppressed in the first model due to $R$-symmetry. An additional $Z_4$ symmetry employed in the first model  makes the color triplets suitably lighter than $M_G$. The issue of gauge coupling unification with intermediate mass scale color triplets is resolved with additional vectorlike 5-plets. The doublets in these 5-plets can naturally attain masses of intermediate scale due to a $Z_4$ symmetry. 
The predicted range of various decay rates and branching fractions is presented as a function of color triplet masses. Comparison with other GUTs is discussed with special emphasis on unflipped $SU(5)$ model and other flipped $SU(5)$ models. The decay channel $K^+ \bar{\nu}$  can play a pivotal role in discriminating various models of flipped $SU(5)$ and other GUTs.
\section*{Acknowledgments}
This work is partially supported by the DOE grant No. DE-SC0013880 (Q.S).


\begin{thebibliography}{99}



\bibitem{Miura:2016krn}
K.~Abe \textit{et al.} [Super-K],
``Search for proton decay via $p \to e^+\pi^0$ and $p \to \mu^+\pi^0$ in 0.31  megaton·years exposure of the Super-K water Cherenkov detector,''
Phys. Rev. D \textbf{95}, no.1, 012004 (2017)
[arXiv:1610.03597 [hep-ex]].




\bibitem{Abe:2013lua}
K.~Abe \textit{et al.} [Super-Kamiokande],
``Search for Nucleon Decay via $n \to \bar{\nu} \pi^{0}$ and $p \to \bar{\nu} \pi^{+}$ in Super-Kamiokande,''
Phys. Rev. Lett. \textbf{113}, no.12, 121802 (2014)
[arXiv:1305.4391 [hep-ex]].





\bibitem{Takhistov:2016eqm}
V.~Takhistov [Super-Kamiokande],
``Review of Nucleon Decay Searches at Super-Kamiokande,''
[arXiv:1605.03235 [hep-ex]].






\bibitem{Regis:2012sn}
C.~Regis \textit{et al.} [Super-Kamiokande],
``Search for Proton Decay via $p \rightarrow \mu^+ K^0$ in Super-Kamiokande I, II, and III,''
Phys. Rev. D \textbf{86}, 012006 (2012)
[arXiv:1205.6538 [hep-ex]].



\bibitem{Abe:2014mwa}
K.~Abe \textit{et al.} [Super-Kamiokande],
``Search for proton decay via $p \rightarrow \nu \nu K^+$ using 260  kiloton year data of Super-Kamiokande,''
Phys. Rev. D \textbf{90}, no.7, 072005 (2014)
[arXiv:1408.1195 [hep-ex]].




\bibitem{Kobayashi:2005pe}
K.~Kobayashi \textit{et al.} [Super-Kamiokande],
``Search for nucleon decay via modes favored by supersymmetric grand unification models in Super-Kamiokande-I,''
Phys. Rev. D \textbf{72}, 052007 (2005)
[arXiv:hep-ex/0502026 [hep-ex]].



\bibitem{An:2015jdp}
F.~An \textit{et al.} [JUNO],
``Neutrino Physics with JUNO,''
J. Phys. G \textbf{43}, no.3, 030401 (2016)
[arXiv:1507.05613 [physics.ins-det]].



\bibitem{Acciarri:2015uup}
R.~Acciarri \textit{et al.} [DUNE],
``Long-Baseline Neutrino Facility (LBNF) and Deep Underground Neutrino Experiment (DUNE): Conceptual Design Report, Volume 2: The Physics Program for DUNE at LBNF,''
[arXiv:1512.06148 [physics.ins-det]];
B.~Abi \textit{et al.} [DUNE],
``Deep Underground Neutrino Experiment (DUNE), Far Detector Technical Design Report, Volume II DUNE Physics,''
[arXiv:2002.03005 [hep-ex]].




\bibitem{Abe:2018uyc}
K.~Abe \textit{et al.} [Hyper-Kamiokande],
``Hyper-Kamiokande Design Report,''
[arXiv:1805.04163 [physics.ins-det]].





\bibitem{Goto:1998qg}
T.~Goto and T.~Nihei,
``Effect of RRRR dimension five operator on the proton decay in the minimal SU(5) SUGRA GUT model,''
Phys. Rev. D \textbf{59}, 115009 (1999)
[arXiv:hep-ph/9808255 [hep-ph]].



\bibitem{Murayama:2001ur}
H.~Murayama and A.~Pierce,
``Not even decoupling can save minimal supersymmetric SU(5),''
Phys. Rev. D \textbf{65}, 055009 (2002)
[arXiv:hep-ph/0108104 [hep-ph]].



\bibitem{Hisano:2013exa}
J.~Hisano, D.~Kobayashi, T.~Kuwahara and N.~Nagata,
``Decoupling Can Revive Minimal Supersymmetric SU(5),''
JHEP \textbf{07}, 038 (2013)
[arXiv:1304.3651 [hep-ph]].


\bibitem{Nagata:2013sba}
N.~Nagata and S.~Shirai,
``Sfermion Flavor and Proton Decay in High-Scale Supersymmetry,''
JHEP \textbf{03}, 049 (2014)
[arXiv:1312.7854 [hep-ph]].




\bibitem{Nagata:2013ive}
N.~Nagata,
``Proton Decay in High-scale Supersymmetry,''


\bibitem{Evans:2015bxa}
J.~L.~Evans, N.~Nagata and K.~A.~Olive,
``$SU(5)$ Grand Unification in Pure Gravity Mediation,''
Phys. Rev. D \textbf{91}, 055027 (2015)
[arXiv:1502.00034 [hep-ph]].



\bibitem{Ellis:2015rya}
J.~Ellis, J.~L.~Evans, F.~Luo, N.~Nagata, K.~A.~Olive and P.~Sandick,
``Beyond the CMSSM without an Accelerator: Proton Decay and Direct Dark Matter Detection,''
Eur. Phys. J. C \textbf{76}, no.1, 8 (2016)
[arXiv:1509.08838 [hep-ph]].



\bibitem{Ellis:2016tjc}
J.~Ellis, J.~L.~Evans, A.~Mustafayev, N.~Nagata and K.~A.~Olive,
``The Super-GUT CMSSM Revisited,''
Eur. Phys. J. C \textbf{76}, no.11, 592 (2016)
[arXiv:1608.05370 [hep-ph]].


\bibitem{Evans:2019oyw}
J.~L.~Evans, N.~Nagata and K.~A.~Olive,
``A Minimal $SU(5)$ SuperGUT in Pure Gravity Mediation,''
Eur. Phys. J. C \textbf{79}, no.6, 490 (2019)
[arXiv:1902.09084 [hep-ph]].



\bibitem{Ellis:2019fwf}
J.~Ellis, J.~L.~Evans, N.~Nagata, K.~A.~Olive and L.~Velasco-Sevilla,
``Supersymmetric proton decay revisited,''
Eur. Phys. J. C \textbf{80}, no.4, 332 (2020)
[arXiv:1912.04888 [hep-ph]].




\bibitem{LHC}
M.~Aaboud {\it et al.} [ATLAS Collaboration],
``Search for supersymmetry in final states with missing transverse momentum and multiple $b$-jets in proton-proton collisions at $ \sqrt{s}=13 $ TeV with the ATLAS detector,''
JHEP {\bf 1806}, 107 (2018)
[arXiv:1711.01901 [hep-ex]];
M.~Aaboud {\it et al.} [ATLAS Collaboration],
``Search for squarks and gluinos in final states with jets and missing transverse momentum using 36 fb$^{-1}$ of $\sqrt{s}=13$ TeV pp collision data with the ATLAS detector,''
Phys.\ Rev.\ D {\bf 97}, no. 11, 112001 (2018)
[arXiv:1712.02332 [hep-ex]];
A.~M.~Sirunyan {\it et al.} [CMS Collaboration],
``Search for new phenomena with the $M_{\mathrm {T2}}$ variable in the all-hadronic final state produced in proton-proton collisions at $\sqrt{s} = 13$ $\,\text {TeV}$,''
Eur.\ Phys.\ J.\ C {\bf 77}, no. 10, 710 (2017)
[arXiv:1705.04650 [hep-ex]];
A.~M.~Sirunyan {\it et al.} [CMS Collaboration],
``Search for natural and split supersymmetry in proton-proton collisions at $ \sqrt{s}=13 $ TeV in final states with jets and missing transverse momentum,''
JHEP {\bf 1805}, 025 (2018)
[arXiv:1802.02110 [hep-ex]].




\bibitem{DeRujula:1980qc}
A.~De Rujula, H.~Georgi and S.~L.~Glashow,
``Flavor Goniometry by Proton Decay,''
Phys. Rev. Lett. \textbf{45}, 413 (1980)


\bibitem{Georgi:1980pw}
H.~Georgi, S.~L.~Glashow and M.~Machacek,
``$\mu^+$ Polarization in Proton Decay: A Probe of Flavor Mixing in Unified Models,''
Phys. Rev. D \textbf{23}, 783 (1981)



\bibitem{Barr:1981qv}
S.~M.~Barr,
``A New Symmetry Breaking Pattern for $SO(10)$ and Proton Decay'',
Phys.\ Lett.\  {\bf 112B} (1982) 219.



\bibitem{Derendinger:1983aj}
J.~P.~Derendinger, J.~E.~Kim and D.~V.~Nanopoulos,
``Anti-$SU(5)$,''
Phys. Lett. B \textbf{139}, 170-176 (1984)


\bibitem{Antoniadis:1987dx}
I.~Antoniadis, J.~R.~Ellis, J.~S.~Hagelin and D.~V.~Nanopoulos,
``Supersymmetric Flipped $SU(5)$ Revitalized,''
Phys. Lett. B \textbf{194}, 231-235 (1987)


	
\bibitem{Barr:1988yj}
S.~M.~Barr,
``Some Comments on Flipped $SU(5) \times U(1)$ and Flipped Unification in General,''
Phys. Rev. D \textbf{40}, 2457 (1989)


\bibitem{Shafi:1998dv}
Q.~Shafi and Z.~Tavartkiladze,
``Atmospheric and solar neutrino oscillations in neutrino $\nu$ MSSM and beyond,''
Phys. Lett. B \textbf{448}, 46-56 (1999)
[erratum: Phys. Lett. B \textbf{450}, 480-480 (1999)]
[arXiv:hep-ph/9811463 [hep-ph]].





\bibitem{Ellis:2020qad}
J.~Ellis, M.~A.~G.~Garcia, N.~Nagata, D.~V.~Nanopoulos and K.~A.~Olive,
``Proton Decay: Flipped vs Unflipped SU(5),''
JHEP \textbf{05}, 021 (2020)
[arXiv:2003.03285 [hep-ph]].



\bibitem{Lazarides:2020bgy}
G.~Lazarides, M.~U.~Rehman and Q.~Shafi,
``Proton Decay in Supersymmetric $SU(4)_c \times SU(2)_L \times SU(2)_R$,''
[arXiv:2007.15317 [hep-ph]].




\bibitem{Lazarides:2020zof}
G.~Lazarides, M.~U.~Rehman, Q.~Shafi and F.~K.~Vardag,
``Shifted $\mu$-hybrid inflation, gravitino dark matter, and observable gravity waves,''
[arXiv:2007.01474 [hep-ph]].




\bibitem{Kyae:2005nv}
B.~Kyae and Q.~Shafi,
``Flipped $SU(5)$ predicts $\delta T/T$,''
Phys. Lett. B \textbf{635}, 247-252 (2006)
[arXiv:hep-ph/0510105 [hep-ph]].



\bibitem{Rehman:2009yj}
M.~U.~Rehman, Q.~Shafi and J.~R.~Wickman,
``Minimal Supersymmetric Hybrid Inflation, Flipped $SU(5)$ and Proton Decay,''
Phys. Lett. B \textbf{688}, 75-81 (2010)
[arXiv:0912.4737 [hep-ph]].



\bibitem{Rehman:2018nsn}
M.~U.~Rehman, Q.~Shafi and U.~Zubair,
``Gravity waves and proton decay in a flipped $SU(5)$ hybrid inflation model,''
Phys. Rev. D \textbf{97}, no.12, 123522 (2018)
[arXiv:1804.02493 [hep-ph]].




\bibitem{Hamaguchi:2020tet}
K.~Hamaguchi, S.~Hor and N.~Nagata,
``R-Symmetric Flipped $SU(5)$,''
[arXiv:2008.08940 [hep-ph]].

\bibitem{Rehman:2018gnr}
M.~U.~Rehman, M.~M.~A.~Abid and A.~Ejaz,
``New Inflation in Supersymmetric $SU(5)$ and Flipped $SU(5)$ GUT Models,''
[arXiv:1804.07619 [hep-ph]].



\bibitem{Giudice:1988yz}
G.~F.~Giudice and A.~Masiero,
``A Natural Solution to the mu Problem in Supergravity Theories,''
Phys. Lett. B \textbf{206}, 480-484 (1988)



\bibitem{Esteban:2020cvm}
I.~Esteban, M.~C.~Gonzalez-Garcia, M.~Maltoni, T.~Schwetz and A.~Zhou,
``The fate of hints: updated global analysis of three-flavor neutrino oscillations,''
[arXiv:2007.14792 [hep-ph]].




\bibitem{Tanabashi:2018oca}
M.~Tanabashi \textit{et al.} [Particle Data Group],
``Review of Particle Physics,''
Phys. Rev. D \textbf{98}, no.3, 030001 (2018)



\bibitem{Mohapatra:1986bd}
R.~N.~Mohapatra and J.~W.~F.~Valle,
``Neutrino Mass and Baryon Number Nonconservation in Superstring Models,''
Phys. Rev. D \textbf{34}, 1642 (1986)


\bibitem{Malinsky:2005bi}
M.~Malinsky, J.~C.~Romao and J.~W.~F.~Valle,
``Novel supersymmetric $SO(10)$ seesaw mechanism,''
Phys. Rev. Lett. \textbf{95}, 161801 (2005)
[arXiv:hep-ph/0506296 [hep-ph]].




\bibitem{Deppisch:2004fa}
F.~Deppisch and J.~W.~F.~Valle,
``Enhanced lepton flavor violation in the supersymmetric inverse seesaw model,''
Phys. Rev. D \textbf{72}, 036001 (2005)
[arXiv:hep-ph/0406040 [hep-ph]].




\bibitem{Ellis:2019jha}
J.~Ellis, M.~A.~G.~Garcia, N.~Nagata, D.~V.~Nanopoulos and K.~A.~Olive,
``Cosmology with a master coupling in flipped $SU(5) \times U(1)$: the $\lambda_6$ universe,''
Phys. Lett. B \textbf{797}, 134864 (2019)
[arXiv:1906.08483 [hep-ph]].



\bibitem{Ellis:2019opr}
J.~Ellis, M.~A.~G.~Garcia, N.~Nagata, D.~V.~Nanopoulos and K.~A.~Olive,
``Superstring-Inspired Particle Cosmology: Inflation, Neutrino Masses, Leptogenesis, Dark Matter \& the SUSY Scale,''
JCAP \textbf{01}, 035 (2020)
[arXiv:1910.11755 [hep-ph]].





\bibitem{Civiletti:2013cra}
M.~Civiletti, M.~Ur Rehman, E.~Sabo, Q.~Shafi and J.~Wickman,
``R-symmetry breaking in supersymmetric hybrid inflation,''
Phys. Rev. D \textbf{88}, no.10, 103514 (2013)
[arXiv:1303.3602 [hep-ph]].




\bibitem{Minkowski:1977sc} 
  P.~Minkowski,
  ``$\mu \rightarrow e \gamma$ at a Rate of One Out of 1-Billion Muon Decays?,''
  Phys.\ Lett.\ B {\bf 67}, 421 (1977);
  T.~Yanagida,
  ``Horizontal Symmetry And Masses Of Neutrinos,''
  Conf.\ Proc.\ C {\bf 7902131}, 95 (1979);
  M.~Gell-Mann, P.~Ramond and R.~Slansky,
  ``Complex Spinors and Unified Theories,''
  Conf.\ Proc.\ C {\bf 790927}, 315 (1979)
  [arXiv:1306.4669 [hep-th]];
  S.~L.~Glashow,
  ``The Future of Elementary Particle Physics,''
  NATO Sci.\ Ser.\ B {\bf 59}, 687 (1980);
  R.~N.~Mohapatra and G.~Senjanovic,
  ``Neutrino Mass and Spontaneous Parity Violation,''
  Phys.\ Rev.\ Lett.\  {\bf 44}, 912 (1980);
  J.~Schechter and J.~W.~F.~Valle,
  ``Neutrino Masses in $SU(2)\times U(1)$ Theories,''
  Phys.\ Rev.\ D {\bf 22}, 2227 (1980).




\bibitem{Senoguz:2003hc}
V.~N.~Senoguz and Q.~Shafi,
``GUT scale inflation, nonthermal leptogenesis, and atmospheric neutrino oscillations,''
Phys. Lett. B \textbf{582} (2004), 6-14
[arXiv:hep-ph/0309134 [hep-ph]].





\bibitem{Barr:1982pk}
S.~M.~Barr and S.~D.~Ellis,
``Proton-decay branching ratios in $SO(10)$,''
Phys. Rev. D \textbf{27}, 1190 (1983)


\bibitem{Ellis:1993ks}
J.~R.~Ellis, J.~L.~Lopez, D.~V.~Nanopoulos and K.~A.~Olive,
``Flipped angles and phases: A Systematic study,''
Phys. Lett. B \textbf{308}, 70-78 (1993)
[arXiv:hep-ph/9303307 [hep-ph]].



\bibitem{Ellis:1995at}
J.~R.~Ellis, J.~L.~Lopez and D.~V.~Nanopoulos,
``Lowering alpha-s by flipping $SU(5)$,''
Phys. Lett. B \textbf{371}, 65-70 (1996)
[arXiv:hep-ph/9510246 [hep-ph]].




\bibitem{Ellis:2002vk}
J.~R.~Ellis, D.~V.~Nanopoulos and J.~Walker,
``Flipping $SU(5)$ out of trouble,''
Phys. Lett. B \textbf{550}, 99-107 (2002)
[arXiv:hep-ph/0205336 [hep-ph]].


\bibitem{Dorsner:2004xx}
I.~Dorsner and P.~Fileviez Perez,
``Distinguishing between $SU(5)$ and flipped $SU(5)$,''
Phys. Lett. B \textbf{605}, 391-398 (2005)
[arXiv:hep-ph/0409095 [hep-ph]].




\bibitem{Li:2010ar}
T.~Li, D.~V.~Nanopoulos and J.~W.~Walker,
``Fast proton decay,''
Phys. Lett. B \textbf{693}, 580-583 (2010)
[arXiv:0910.0860 [hep-ph]];
T.~Li, D.~V.~Nanopoulos and J.~W.~Walker,
``Elements of F-ast Proton Decay,''
Nucl. Phys. B \textbf{846}, 43-99 (2011)
[arXiv:1003.2570 [hep-ph]].





\bibitem{Dvali:1997uq}
G.~R.~Dvali, G.~Lazarides and Q.~Shafi,
``Mu problem and hybrid inflation in supersymmetric $SU(2)_L \times SU(2)_R \times U(1)_{(B-L)}$,''
Phys. Lett. B \textbf{424}, 259-264 (1998)
[arXiv:hep-ph/9710314 [hep-ph]].










\bibitem{Hisano:2013ege}
J.~Hisano, D.~Kobayashi, Y.~Muramatsu and N.~Nagata,
``Two-loop Renormalization Factors of Dimension-six Proton Decay Operators in the Supersymmetric Standard Models,''
Phys. Lett. B \textbf{724}, 283-287 (2013)
[arXiv:1302.2194 [hep-ph]].



\bibitem{Abbott:1980zj}
L.~F.~Abbott and M.~B.~Wise,
``The Effective Hamiltonian for Nucleon Decay,''
Phys. Rev. D \textbf{22}, 2208 (1980)


\bibitem{Munoz:1986kq}
C.~Munoz,
``Enhancement Factors for Supersymmetric Proton Decay in $SU(5)$ and $SO(10)$ With Superfield Techniques,''
Phys. Lett. B \textbf{177}, 55-59 (1986)


\bibitem{Nihei:1994tx}
T.~Nihei and J.~Arafune,
``The Two loop long range effect on the proton decay effective Lagrangian,''
Prog. Theor. Phys. \textbf{93}, 665-669 (1995)
[arXiv:hep-ph/9412325 [hep-ph]].


\bibitem{Aoki:2017puj}
Y.~Aoki, T.~Izubuchi, E.~Shintani and A.~Soni,
``Improved lattice computation of proton decay matrix elements,''
Phys. Rev. D \textbf{96}, no.1, 014506 (2017)
[arXiv:1705.01338 [hep-lat]].




\bibitem{Babu:1997js}
K.~S.~Babu, J.~C.~Pati and F.~Wilczek,
``Suggested new modes in supersymmetric proton decay,''
Phys. Lett. B \textbf{423}, 337-347 (1998)
[arXiv:hep-ph/9712307 [hep-ph]].



\bibitem{Babu:1998wi}
K.~S.~Babu, J.~C.~Pati and F.~Wilczek,
``Fermion masses, neutrino oscillations, and proton decay in the light of Super-Kamiokande,''
Nucl. Phys. B \textbf{566}, 33-91 (2000)
[arXiv:hep-ph/9812538 [hep-ph]].





\bibitem{Haba:2020bls}
N.~Haba, Y.~Mimura and T.~Yamada,
``Enhanced $\Gamma(p\to K^0\mu^+)/\Gamma(p\to K^+\bar{\nu}_\mu)$ as a Signature of Minimal Renormalizable SUSY $SO(10)$ GUT,''
[arXiv:2002.11413 [hep-ph]].



\bibitem{Rehman:2012gd}
M.~U.~Rehman and Q.~Shafi,
``Simplified Smooth Inflation with Observable Gravity Waves,''
Phys. Rev. D \textbf{86} (2012), 027301
[arXiv:1202.0011 [hep-ph]].






\bibitem{Pati:2002ig}
J.~C.~Pati,
``Confronting the conventional ideas of grand unification with fermion masses, neutrino oscillations and proton decay,''
ICTP Lect. Notes Ser. \textbf{10}, 113-182 (2002)
[arXiv:hep-ph/0204240 [hep-ph]].




\end{thebibliography}
\end{document}